\documentclass[pra,aps,showpacs]{revtex4}
\usepackage{epsfig}
\usepackage{amsmath,amssymb,amsthm}
\usepackage{graphicx}

\newcommand{\la}{\langle}
\newcommand{\ra}{\rangle}
\newcommand{\kDE}{l(t)}
\newcommand{\au}{\alpha_u(t)}
\newcommand{\dau}{{\dot\alpha}_u(t)}

\begin{document}
\title{Solving satisfiability problems by fluctuations: The dynamics
  of stochastic local search algorithms}

\author{Wolfgang Barthel, Alexander K. Hartmann, and Martin Weigt} 

\affiliation{Institut f\"ur Theoretische Physik, Universit\"at
  G\"ottingen, Bunsenstr. 9, D-37073 G\"ottingen, Germany}

\date{\today}

\begin{abstract}
  Stochastic local search algorithms are frequently used to
  numerically solve hard combinatorial optimization or decision
  problems. We give numerical and approximate analytical descriptions
  of the dynamics of such algorithms applied to random satisfiability
  problems. We find two different dynamical regimes, depending on the
  number of constraints per variable: For low constraintness, the
  problems are solved efficiently, i.e. in linear time. For higher
  constraintness, the solution times become exponential. We observe
  that the dynamical behavior is characterized by a fast equilibration
  and fluctuations around this equilibrium. If the algorithm runs long
  enough, an exponentially rare fluctuation towards a solution
  appears. 
\end{abstract}


\maketitle

\section{Introduction}
\label{sec:intro}

The last years have seen a fruitful exchange between theoretical
computer science and statistical mechanics \cite{AI,TCS}. Due to the
formal analogy between various combinatorial optimization problems
and certain spin-glass models, substantial progress in the
understanding of hard combinatorial questions could be made by using
tools which were originally developed in the statistical mechanics of
disordered systems.

The most striking results so far were obtained in the description of
the solution-space structure of the random satisfiability problem
\cite{MoZe,nature,BiMoWe,MeZePa,MeZe}, of the number partitioning
problem \cite{Me1,Me2}, of vertex covers \cite{WeHa1,HaWe,WeHa2} or
colorings \cite{MuPaWeZe} of random graphs. In these cases,
equilibrium methods from statistical mechanics can be applied
directly, including e.g.  the replica and cavity approaches. The main
result is that these models undergo phase transitions from an easily
solvable, under-constrained phase to a hard, highly constrained one.
The latter is characterized by the existence of glass-like states,
i.e. the solution space is subdivided into a large number of
disconnected clusters, and there are exponentially many excited states
hindering even the best local algorithms from finding optimal
solutions in sub-exponential time (where exponential means, here and
in the following, exponential in the system size, as given e.g. by the
number of discrete degrees of freedom or, in a more computer-science
oriented language, in the number of bits needed to encode an instance
of the problem under consideration).

Up to now, much less is understood about the dynamical behavior of
algorithms which are used to numerically solve the combinatorial
problems. Also these are known to undergo algorithm-dependent phase
transitions from phase space regions where the problems are typically
efficiently solvable, to regions where solutions are exponentially
hard to construct. Some understanding was obtained for heuristics, i.e.
approximate algorithms running in linear time, see e.g.
\cite{Ac,Fr,We}, for complete solvers \cite{MoCo,WeHa3,MoCo2} which
are guaranteed to find an optimal solution, and finally for randomized
versions of these complete algorithms \cite{MoZe3,MoCo3}. The problem
in analyzing algorithms is that they are intimately related to
non-equilibrium statistical mechanics, which frequently is technically
much harder to handle. In addition, algorithms are not forced to
fulfill physical criteria like detailed balance, which again
complicates the analysis.

In this paper, we are going to analyze a different class of algorithms:
{\it stochastic local search algorithms}, in particular variants of
the so-called {\it walk-SAT} algorithm \cite{SeLeKa} which is one of
the most popular and successful solvers for satisfiability problems.
Whereas the full problem is to hard to attack successfully  by
means of analytical tools, we will give some approximation methods
which allow us to draw a qualitative picture on how these algorithms
solve an optimization problem.

The paper is organized as follows: In Sec. \ref{sec:model} the
considered models are introduced. We first introduce the random
$K$-satisfiability problem ($K$-SAT) and give an overview of the
current state of knowledge. Then we introduce a second model, the
random $K$-XOR-satisfiability problem ($K$-XOR-SAT). Being in many
aspects similar to the $K$-SAT, it has recently attracted some
interest due to its better analytical tractability. In the last part
of Sec. \ref{sec:model} we give a short introduction to some
stochastic local search algorithms, in particular to the famous
walk-SAT algorithm which will be analyzed in the present paper.  We
then show some numerical observations in Sec. \ref{sec:num}.  These
are analytically explained in Secs. \ref{sec:lin} and \ref{sec:exp}.
The first of these two sections deals with the linear-time behavior,
whereas the second one describes the exponential time behavior. Our
results are summarized in the last section.

\emph{Note:} While preparing this paper, we noticed that a
complementary study of the walk-SAT algorithm was carried out
independently by G. Semerjian and R. Monasson \cite{MoSe}.

\section{The models}
\label{sec:model}

\subsection{Random $K$-satisfiability}
\label{sec:ksat}

A random $K$-satisfiability ($K$-SAT) formula $F$ consists of $M$
logical clauses $\{C_\mu\}_{\mu=1,...,M}$ which are defined over a set
of $N$ Boolean variables $\{x_i =0,1\}_{i=1,...,N}$ which can take the
values 0=FALSE and 1=TRUE. Every clause contains $K$ randomly chosen
Boolean variables which are connected by logical OR operations
($\vee$) and appear negated with probability 1/2, e.g.\ $C_\mu=( x_i
\vee \overline{x}_j \vee x_k)$ for $K = 3$. Because of the
OR-conjunction a $K$-SAT-clause is satisfied if at least one of the
$K$ variables has the correct assignment. In the formula $F$ all
clauses are connected by logical AND operations ($\wedge$),
\begin{equation}
  \label{eq:formula}
  F=\bigwedge_{\mu=1}^M C_\mu\ 
\end{equation}
so all clauses have to be satisfied simultaneously in order to satisfy
the formula. For $K=2$, i.e. if each clause connects only two
variables, the problem is easy, and polynomial-time algorithms are
known \cite{GaJo}. On the other hand, the problem becomes
${\rm NP}$-complete for all $K>3$ \cite{GaJo}, so one expects that no
efficient algorithm to solve generic $K$-SAT formulas in
polynomial time can be found.

The considerable attention attracted by the random $K$-SAT problem was
initiated about one decade ago, when the model was numerically
observed \cite{SeKi} to undergo a characteristic phase transition
which is parametrized by the clause-to-variable ratio $\alpha=M/N$.
For $\alpha<4.26$ and sufficiently large system sizes $N$, almost all
3-SAT formulas were found to be satisfiable. For $\alpha>4.26$ this
behavior changes drastically; the formulas are found to be
unsatisfiable with a probability approaching one in the thermodynamic
limit $N\to\infty$. Even more interestingly, this transition was
observed to coincide with a strong exponential peak in the algorithmic
solution time of complete algorithms. The hardest to solve formulas
are thus located close to the phase boundary, and are said to be
critically constrained. 

The observation of this phase transition finally led to the
application of analytical tools developed in the statistical mechanics
of disordered systems, since random $K$-SAT can be mapped to a
spin-glass model on a random hyper-graph. After the pioneering work by
Monasson and Zecchina \cite{MoZe} providing the first analytical
approximation to the phase transition using the replica method, many
efforts were done to improve the analytical understanding. In Ref.
\cite{BiMoWe}, on the basis of a variational approach, a second phase
transition was suggested to appear inside the satisfiable phase: For
very low $\alpha$, the set of all solutions to a $K$-SAT formula was
found to be unstructured, with the exponentially large number of solutions
collected in one large connected cluster. For larger $\alpha$ the
solution space breaks into an exponential number of clusters. Using
the cavity approach, the (probably) exact location of this transition
was established recently for $K=3$. It is given by
$\alpha_d=3.92$ \cite{MeZePa,MeZe}.

\subsection{A simpler but similar model: Random $K$-XOR-SAT}
\label{sec:xorsat}

A model showing a very similar behavior, but being analytically much
more tractable, is given by the random $K$-XOR-SAT problem (in the
physical literature initially denoted as $K$-hSAT \cite{RiWeZe}). The
difference to $K$-SAT is that the variables appearing in the clauses
are connected by logical XOR operations ($\oplus$) instead of OR. A
clause is thus satisfied if an odd number of variables is assigned
correctly, i.e. to TRUE if the variable appears non-negated, and to FALSE if
it appears negated.

The $\oplus$-operation is equivalent to an integer addition modulo 2.
Using this equivalence we can map each clause to a linear equation
(modulo 2), and the formula consequently to a coupled set of $M$
linear equations. The solution of this system can be easily found in
$O(N^3)$ steps. Hence XOR-SAT formulas can be solved efficiently by a
\emph{global} algorithm, i.e. by exploiting the global information
about the instance and its structure in every step. If we use,
however, \emph{local} algorithms like the ones used also for $K$-SAT,
we observe a very similar behavior of both models.

Again, the system can be conveniently parametrized by $\alpha=M/N$.
The numbers given below are valid for $K=3$, but the qualitative
picture is valid for any $K>3$. For $\alpha<0.818$, the formula is
typically easy to solve, the solution space consisting of one large
cluster. In the region $0.818<\alpha<0.918$, the formulas are still
satisfiable with probability tending to one for $N\to\infty$, but the
solution state decays into an exponential number of clusters. In
addition, there are also exponentially many metastable states which
prevent even the best local algorithms from fast convergence to a
solution. For $\alpha>0.918$, the system is almost surely
unsatisfiable. 

These values were originally calculated using the replica method which
is believed to be exact, but still lacks a rigorous foundation. A very
beautiful result for $K$-XOR-SAT was recently obtained in two
independent works \cite{CoMoDuMa,MeRiZe}: The results given above,
including the ones obtained by one-step replica symmetry broken
calculations, were reproduced using mathematically rigorous methods.

The $K$-XOR-SAT problem is also interesting from a physical point of
view, because it is equivalent to a diluted $K$-spin model. Such
models are frequently discussed in connection to the glass transition,
see e.g. \cite{FrMeRiWeZe}.


\subsection{Stochastic local search algorithms}
\label{sec:sls}

As already mentioned in the introduction, here we are not interested
in the solution space structure of $K$-SAT and $K$-XOR-SAT, but in the
{\it non-equilibrium dynamics} of so-called stochastic local search
algorithms (SLS).

The idea behind these algorithms is that, if a formula is satisfiable,
a solution can frequently be found more quickly if {\it randomized}
algorithms are used. In general, these algorithms are {\it
  incomplete}, i.e.  they stop once they have found a solution, but
they are not guaranteed to really find one. Due to their random
character, they are also not able to prove the unsatisfiability of a
formula. In the case where there is no solution the algorithm just runs for
ever, or until some running-time cutoff is reached.

Here we mainly concentrate on the {\it walk-SAT algorithm} introduced
in \cite{SeLeKa}. Its most recent implementations are available in the
SATLIB \cite{satlib}, and they are one of the best stochastic local
search algorithms for random $K$-SAT. The algorithm starts with a
random assignment to all $N$ variables. Within this assignment, there
is a number $\alpha_s N$ of satisfied clauses, whereas the other
$\alpha_u N = \left(\alpha - \alpha_s\right) N$ are unsatisfied.

In every step, the algorithm selects an unsatisfied clause $C$ randomly
and then one of its $K$ variables $v^*$\\[-6mm]
\begin{itemize}
\item with probability $q$  randomly (\emph{walk}-step),\\[-6mm]
\item with probability $1-q$ the variable in $C$ occurring in the least
  number of satisfied clauses. (\emph{greedy}-step).\\[-6mm]
\end{itemize}
The current assignment of $v^*$ is inverted. All clauses containing
$v^*$ that were unsatisfied before become now satisfied. Clauses that
were satisfied behave differently for the two models under
consideration: For $K$-SAT, a previously satisfied clause becomes
unsatisfied iff $v^*$ was the only correctly assigned variable in this
clause. For $K$-XOR-SAT, every previously satisfied clause containing $v^*$
becomes unsatisfied.

These steps are repeated until no unsatisfied clause is left.  Then
the algorithm has found a solution of formula $F$ and stops. As
noted earlier the algorithm will run for ever if no solution exists.

There are variants for the greedy step: The algorithm could also
select the variable in $C$ leading to the minimal number of
unsatisfied clauses (``maximal gain''), or the one minimizing the
number of previously satisfied clauses which become unsatisfied
(``minimal negative gain''). The second case is equivalent to our
choice for $K$-XOR-SAT. For $K$-SAT they are different due to the fact
that not all satisfied clauses become unsatisfied.

A completely different heuristic is the GSAT heuristic \cite{SeLeMi}
which, in the greedy step, globally selects the variable leading to
the minimal number of unsatisfied clauses. In numerical studies, this
selection is outperformed by walk-SAT \cite{HoSt}. There also other
heuristic variations of walk-SAT and GSAT are discussed. For reasons
of clarity we concentrate completely on the algorithm given above. We
expect, however, that the approximate approach developed in this paper
can also be extended to more involved cases, as long as the dynamics
can be considered as a Markov process.

A different iteration of variable flips was introduced by Sch\"oning
\cite{Sc}. He suggested to stop the algorithm after $3N$ steps, and to
restart it by selecting a new random initial assignment to all $N$
Boolean variables. For $q=1$, i.e. for a pure random walk dynamic, he
was able to prove that the worst case solution time goes down from
$2^N$ iterations to only $(4/3)^N$ steps, i.e. the algorithm is
exponentially accelerated. This simple algorithm shows, up to a
refinement leading to $1.3303^N$ steps \cite{Sc2}, the currently best
known worst case behavior of all SAT-algorithms.

In the following sections, we will analyze both models for
exponential waiting times and for an exponential number of random
restarts. We will concentrate on formulas which are satisfiable, i.e.
on variables-to-clauses ratios inside the satisfiable phase of the
model under consideration. In the unsatisfiable phase there are no
solutions, thus the algorithm cannot terminate by construction.

\section{Numerical results on the behavior of walk-SAT}
\label{sec:num}

Now we  present some numerical observations on the behavior of
walk-SAT applied to randomly generated satisfiability formulas. We
look to $K$-SAT as well as to $K$-XOR-SAT, and we mainly concentrate
on the solution times needed by walk-SAT, and the dynamical evolution
of the number of unsatisfied clauses while the algorithm is running.
Explaining these observations will be the final aim in the following
sections.

\subsection{Random $K$-SAT}
\label{sec:num-ksat}

\begin{figure}[htbp]
  \begin{center}
    \includegraphics[clip,height = 7cm]{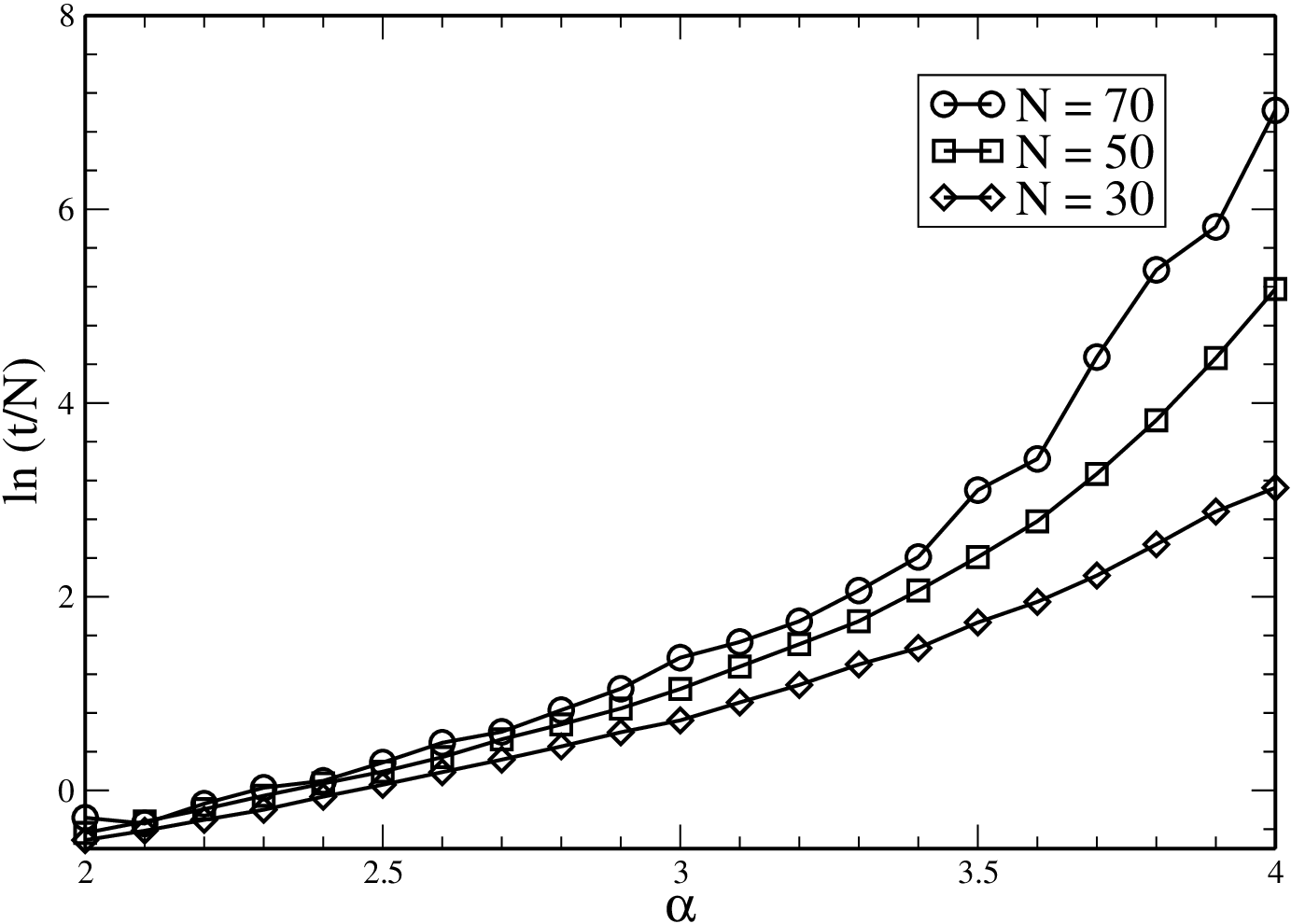}    
\caption{3-SAT: Dependency of the running time of  walk-SAT
  without restarts on the ratio $\alpha$
  of clauses to variables.
    \label{fig:3sat_median}
}
  \end{center}
\end{figure}

\begin{center}
\begin{figure}[htpb]
\includegraphics[clip,height = 7cm]{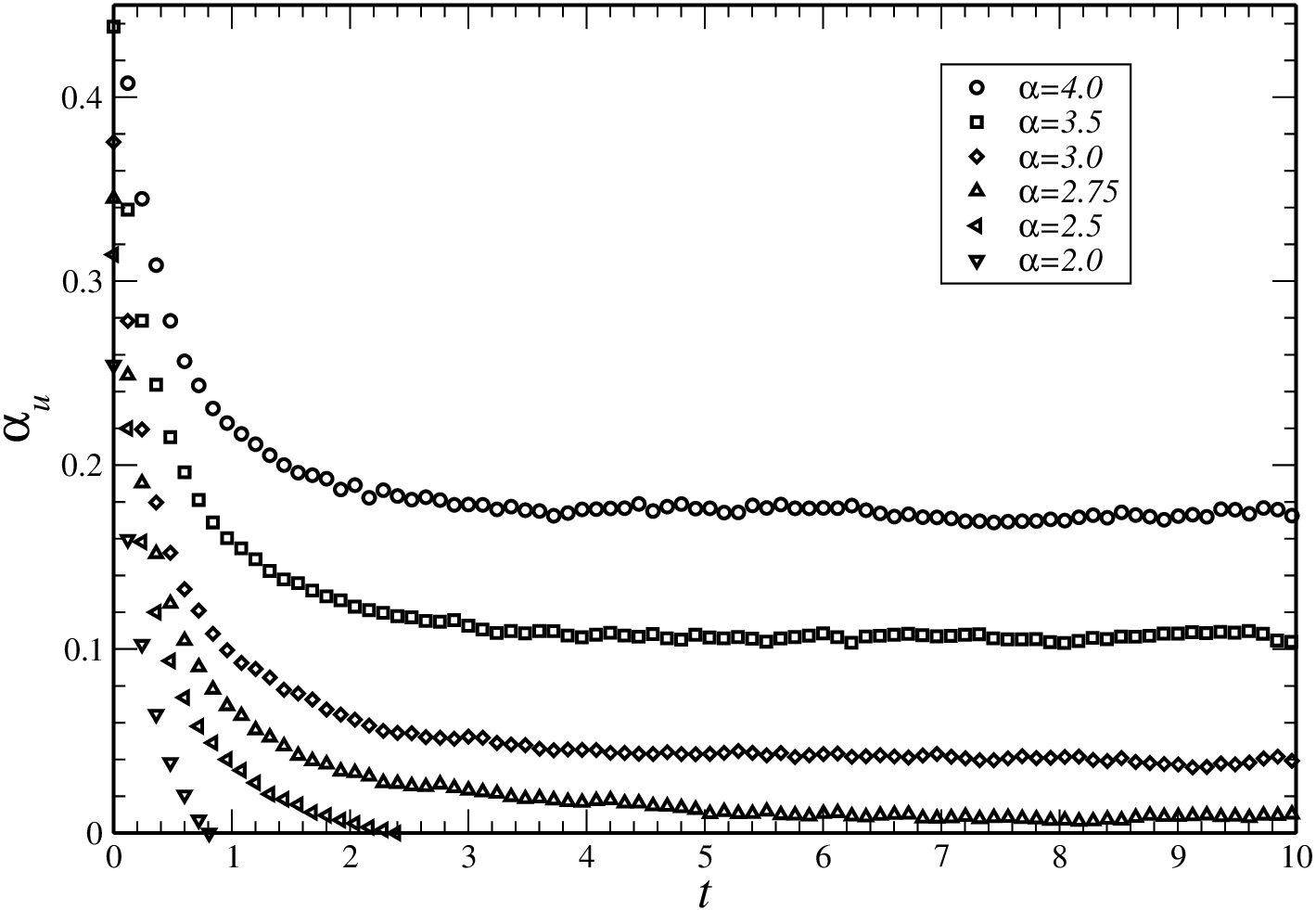}    
\caption{3-SAT: Average number $\alpha_u$ of unsatisfied clauses per
  variable with sample size $N=50000$. Initially this energy density
  quickly decreases. For $\alpha<\alpha_d\simeq 2.7$ it becomes zero
  after a finite time, for larger $\alpha$ a non-zero plateau is
  reached.}
    \label{fig:walksattime}
  \end{figure}
  \end{center}

\begin{figure}[htbp]
  \begin{center}
    \includegraphics[clip, height = 7cm]{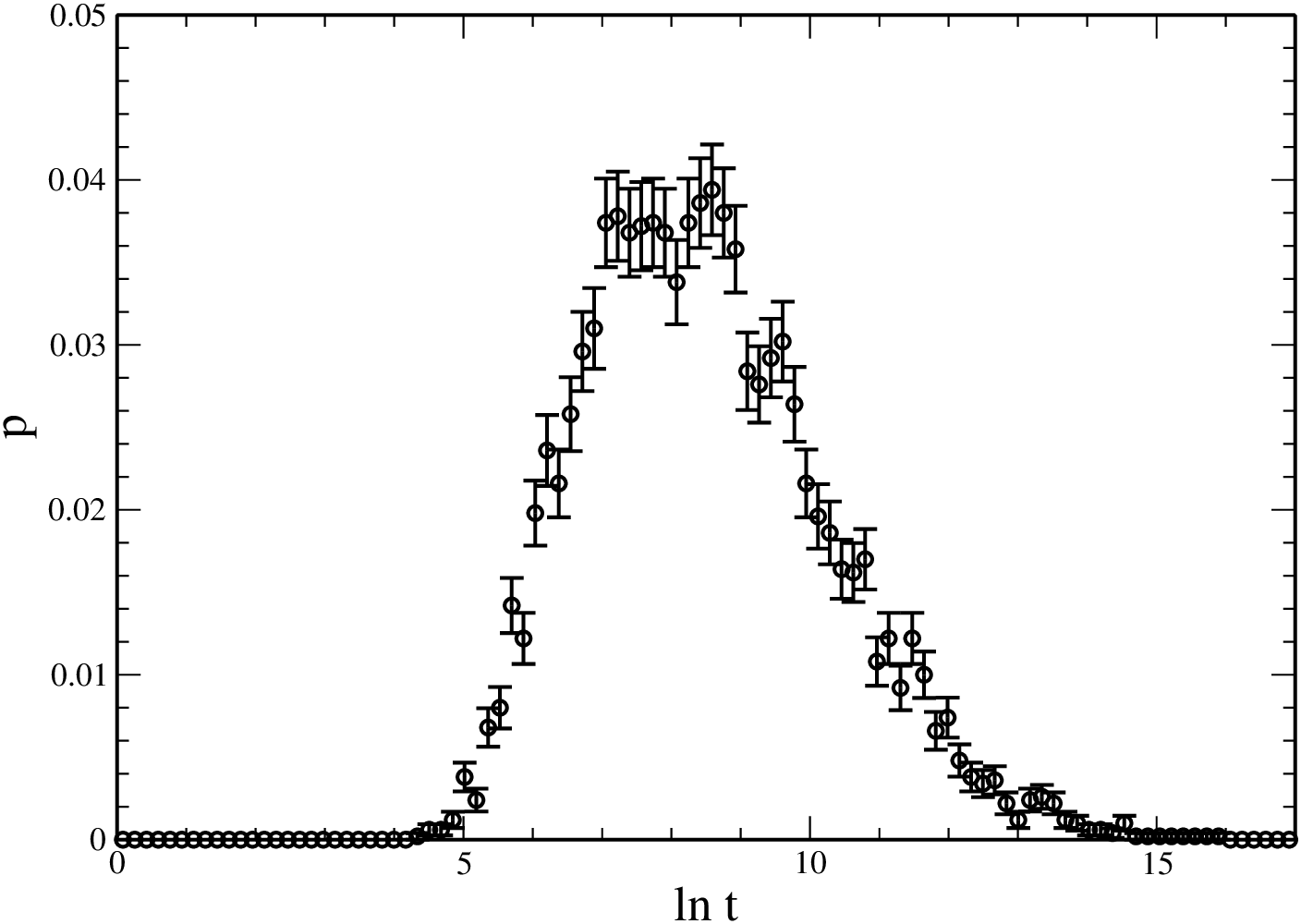}    
\caption{3-SAT: Histogram of the logarithm of the running times of
  walk-SAT without restarts for $\alpha = 3.5$ and $N = 100$.}
    \label{fig:3sat_timehisto}
  \end{center}
\end{figure}

Let us start with random $K$-SAT. At first we realize that the running
time heavily depends on the ratio $\alpha = M/N$ of clauses to
variables.  Let us concentrate on the case $K=3$ and $q = 1$ first,
i.e. only walk-steps are performed. For small $\alpha$ negating one
variable in an unsatisfied clause rarely causes other clauses to
become unsatisfied. Up to a critical threshold $\alpha_d \simeq 2.7$ a
solution is found in a median time growing linearly with $N$, above
this point running times grow exponentially, see Figs.
\ref{fig:3sat_median} and \ref{fig:walksattime}. This observation does
not depend on the fact whether we use the algorithm with or without
restarts. In the following we measure all running times in the number
of Monte-Carlo sweeps (MC sweeps), i.e. a single step of the algorithm
leading to the negation of one variable is counted as $\Delta t =
1/N$. During a time interval of length one, every variable becomes
thus negated on average once.  Note that in this representation linear
solution times lead to a constant number of MC sweeps, whereas
exponential iterations of walk-SAT correspond to exponentially many MC
sweeps. In Fig.  \ref{fig:3sat_timehisto} we show a histogram of the
resolution times inside the exponential regime. Obviously, this
distribution can be well described by the mean of the logarithm of
the running time. For such an exponentially dominated distribution this
is equivalent to characterizing it  by the median, 
 whereas the average running time would be dominated
by exponentially rare events with exponentially longer resolution
times. 

\begin{figure}[htpb]
  \begin{center}
    \includegraphics[clip,height = 7cm]{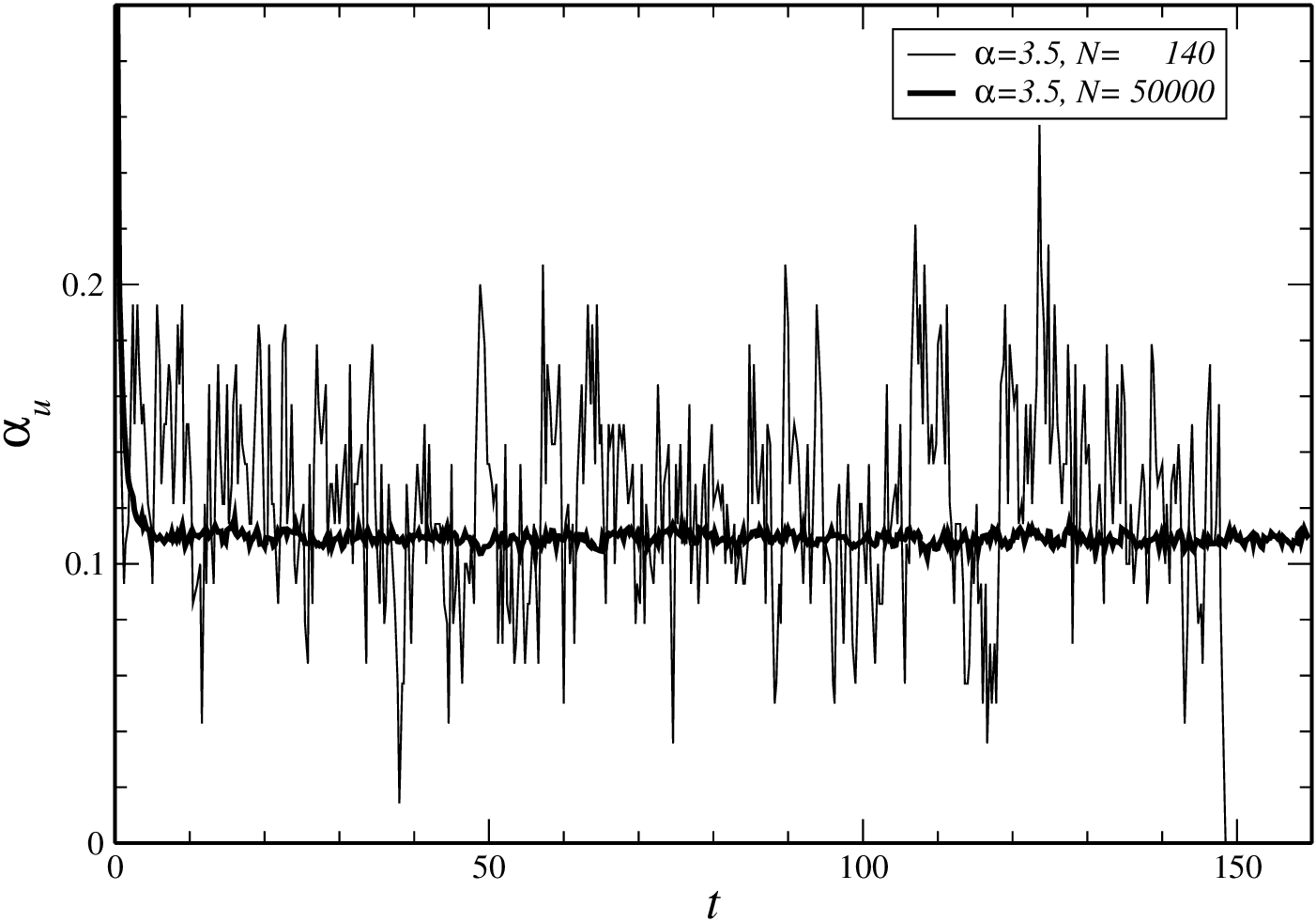}
\caption{After the initial decrease $\alpha_u$ fluctuates around its
  plateau value. Two different system sizes are shown. For the smaller
  one with $N = 150$ a fluctuation after about 145 MC-Steps was large
  enough to reach a solution of the formula.}
    \label{fig:walksattimefluct}
  \end{center}
  \end{figure}

The algorithm starts with an extensive number of unsatisfied
clauses, and stops when their number reaches zero.  To characterize
the search process we therefore look at the behavior of
$\alpha_u(t)$, which is given as the number of unsatisfied clauses per
variable. We can think of it as an energy density of the system of the
$N$ variables. In a randomly drawn starting configuration of the
Boolean variables $x_i,\ i=1,...,N$, there are on average $1/8$ of all
clauses unsatisfied, we thus have almost surely $\alpha_u(t=0) =
\frac{M/8}{N} = \alpha / 8$. Concentrating first on the linear
time behavior, i.e. to finite MC times, it is convenient to work with
large systems, $N\gg 1$. These show a good separation of linear and
exponential time scales but also minimize the influence of
fluctuations. Numerically we find, in dependence on $\alpha$, the
following behavior:\\[-6mm]
\begin{itemize}
\item For $\alpha < \alpha_d$ a solution is found after a finite
  number of MC sweeps, i.e. $\alpha_u(t)$ becomes zero  at finite
  MC times. This solution time grows with $\alpha$, and diverges once
  we approach the dynamical threshold $\alpha_d$.\\[-6mm]
\item For $\alpha > \alpha_d$ the energy density $\alpha_u(t)$
  initially decreases and quickly equilibrates to a non-zero plateau
  (Fig.  \ref{fig:walksattime}). For larger times $\alpha_u(t)$
  fluctuates around its plateau value, as can be seen for smaller
  system sizes, cf. Fig \ref{fig:walksattimefluct}. Eventually, and
  only if the formula is satisfiable, one of these fluctuations is
  large enough to reach $\alpha_u(t)=0$.
\end{itemize}

This behavior explains the origin of the title of the paper: For
$\alpha>\alpha_d$, the system equilibrates to a non-zero number of
unsatisfied clauses, and only fluctuations around this equilibrium 
lead the dynamics to satisfying assignments, and the algorithm
stops. Such macroscopic fluctuations appear, of course, only with
exponentially small probability, giving rise to exponential solution
times.

\begin{figure}[htbp]
  \begin{center}
    \includegraphics[clip,height = 7cm]{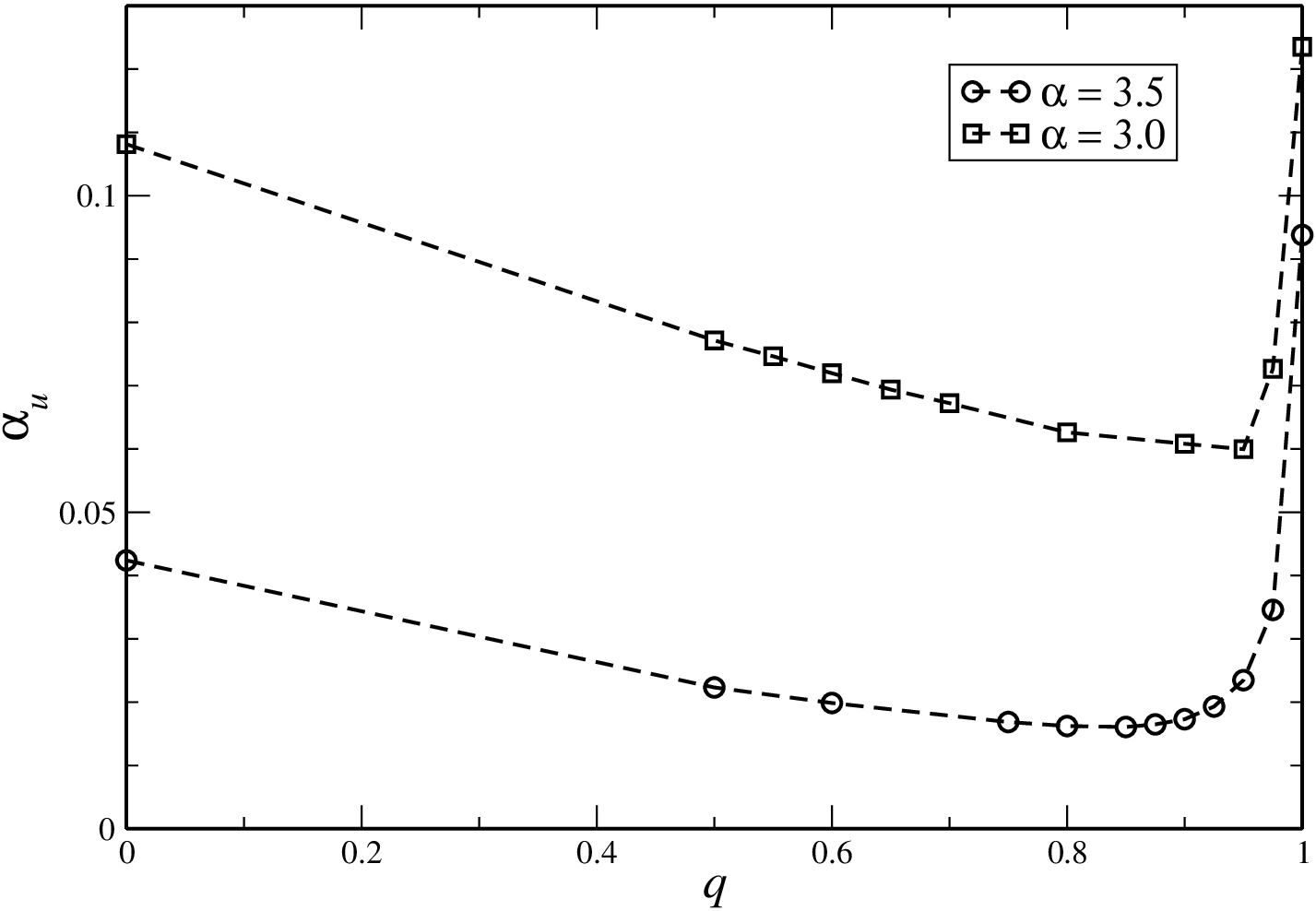}    
\caption{3-SAT: Plateau energy for $\alpha = 3.5$ and $\alpha=3.0$
  depending on the fraction $q$ of greedy steps performed by the
  algorithm. The plateau energy is minimal for $q=0.95$ resp. $q=0.85$.}
    \label{fig:walksattimegreedy}
  \end{center}
\end{figure}

This observation leads to an obvious way of improving the algorithmic
performance: We may choose a better heuristic having a lower
equilibrium number of unsatisfied clauses. Exactly this is achieved by
introducing a fraction $q>0$ of greedy steps, see Fig.
\ref{fig:walksattimegreedy} where the plateau energy is determined as
a function of $q$ for two different values of $\alpha>\alpha_d$. We
can see an minimum in the plateau energy for high values of $q$. The
dynamical threshold itself also changes slightly and has a maximum at
$q\simeq 0.85$. There formulas up to $\alpha \simeq 2.8$ can be solved
in linear time.

\subsection{Random $K$-XOR-SAT}
\label{sec:num-kxorsat}

A qualitatively similar behavior can be observed for random
$K$-XOR-SAT, for $K=3$ and $q=1$ (pure walk dynamics). The main
difference is of a quantitative nature: the dynamical threshold marking
the onset of exponential solution times is located at $\alpha_d \simeq
0.33$. We therefore do not repeat the figures given for random 3-SAT,
but the corresponding numerical data can be found in the following
sections in comparison to analytical results.




\section{A rate-equation approach to the linear-time behavior}
\label{sec:lin}

The main idea of the analytical approach presented in this
section is to characterize each variable only by the number of
satisfied and unsatisfied clauses it is contained in. We subdivide the
set of all $N$ Boolean variables into subsets of $N_t(s,u)$ variables
belonging to $s$ satisfied and $u$ unsatisfied clauses, for a randomly
selected variable the numbers $s$ and $u$ are thus taken with
probability $p_t(s,u)= N_t(s,u) / N$. The numbers $N_t(s,u)$ and thus
also the probabilities $p_t(s,u)$ are changed by the action of
walk-SAT, but for every single variable $s+u$ remains constant as it
counts the total number of clauses containing this variable.

From these quantities we can, in particular, calculate the total
number of unsatisfied clauses $N \alpha_u(t) $. Taking into account
that by summing over variables every clause is counted $K$-fold, we
find
\begin{equation}
  \label{eq:alphaupsu}
  \alpha_u(t) = \frac{\la u \ra_t}K
\end{equation}
where $\la\cdot\ra_t = \sum_{s,u}(\cdot) p_t(s, u)$ denotes the
average over the distribution $p_t$ at MC-time $t$.

The algorithm does not select variables according to $p_t(s,u)$, but
selects first an unsatisfied clause $C^*$ and then, according to the
chosen heuristic (greedy or walk step), one of the variables $v^*$ in
$C^*$ is flipped. The probability that variable $v^*$ belongs to
exactly $s$ satisfied and $u$ unsatisfied clauses is denoted by
$p^{(flip)}_t(s,u)$, and can be calculated from $p_t(s,u)$ under the
assumption of {\it independence of neighboring sites}, i.e. we assume
that the joint distribution for three variables being in one
unsatisfied clause factorizes. This assumption, which we will exploit
more frequently, is the main approximation we apply in the analytical
approach, and it allows us to describe the full dynamics in terms of
$p_t(s,u)$. It is strictly valid only for the initial configuration of
the dynamics, but as we will see below, it can give a good
approximation also for larger times.

For a walk step, variable $v^*$ is randomly selected in $C^*$. There
are $u N_t(s,u)$ possibilities for selecting a $v^*$ which appears in 
$s$ satisfied and $u$ unsatisfied clauses. By normalization we thus
find the following selection probability:
\begin{equation}
  \label{eq:p_flip_walk}
  p^{(flip-walk)}_t(s, u) =  \frac{u p_{t}(s,u)}{\la u \ra_t}
  =: p^{(u)}_t(s, u)\ .
\end{equation}

For a greedy step, the only random choice is the selection of the
unsatisfied clause $C^*$. Then the variable $v^*$ is selected which
appears in the smallest number $s$ of satisfied clauses among all $K$
variables in $C^*$. If there is more than one variable with the
same minimal $s$, then one of them is chosen randomly. Applying the
independent-site assumption, and using the Heavyside function under
the convention $\Theta(0) = 1/2$, we find for $K=2$
\begin{eqnarray}
  \label{eq:pflip_greedy_2}
  p^{(flip-2-greedy)}_t(s, u) &=& \sum_{s_1, u_1, s_2, u_2}
  p_t^{(u)}(s_1,u_1) p_t^{(u)}(s_2,u_2)
  \left[ \delta_{(s_1,u_1),(s,u)}\cdot \Theta(s_2-s_1)
    +\delta_{(s_2,u_2),(s,u)}\cdot \Theta(s_1-s_2)
    \right]\notag\\
    &=& 2 p_t^{(u)}(s, u) \sum_{s', u'} p_t^{(u)}(s', u') \Theta(s'-s)\notag\\
    &=& p_t^{(u)}(s, u) \left[2 - \sum_{u'=0}^{\infty}
      \left(p_t^{(u)}(s, u') + 2 \sum_{s'=0}^{s-1} p_t^{(u)}(s',
        u')\right)\right]\ ,
\end{eqnarray}
and similarly for $K=3$
\begin{eqnarray}
  \label{eq:pflip_greedy_3}
  p^{(flip-3-greedy)}_t(s, u) 
    &=& 3 p_t^{(u)}(s, u) \sum_{s', u', s'', u''} p_t^{(u)}(s',
      u') p_t^{(u)}(s'', u'') 
      \left[\Theta(s'-s)\Theta(s''-s)+1/12 \delta_{s,s'}
        \delta_{s,s''}\right]\notag\\ 
    &=& 3 p_t^{(u)}(s, u) \left[1 - \sum_{u'=0}^{\infty}
      \left(1/2 p_t^{(u)}(s, u') +  \sum_{s'=0}^{s-1} p_t^{(u)}(s', 
        u')\right)\right]^2
    +1/4 p_t^{(u)}(s, u) \left[ \sum_{u'=0}^{\infty} p_t^{(u)}(s,
      u')\right]^2\ . 
\end{eqnarray}
Note that the contribution $\delta_{s,s'} \delta_{s,s''}/12$ is a
correction term for the case that $s=s'=s''$ which results from the
convention $\Theta(0) = 1/2$.

For the full algorithms, these two different steps appear with
probabilities $q$ and $1-q$. The selection probability
$p^{(flip)}_t(s, u)$ is thus given by the linear combination of the
two cases,
\begin{equation}
  \label{eq:pflip_greedy_k_3}
  p^{(flip)}_t(s, u) = q p^{(flip-walk)}_t(s, u)
  + (1-q) p^{(flip-K-greedy)}_t(s, u)\ .
\end{equation}

At this point, GSAT-like heuristics could also be included, e.g. by
taking $p^{(flip)}_t(s,u)\sim u^\gamma p_t(s,u)$ with $\gamma>1$. This
would guarantee a preferential selection of variables belonging to a
high number of unsatisfied clauses. Here we do not consider this
additional possibility.

\subsection{A Poissonian estimate for the pure walk dynamics}
\label{sec:treshold_estimate}

For a moment we concentrate on the simplified case where the
algorithms uses only walk steps, i.e. to $q=1$ \cite{note}.  We
further assume that $s$ and $u$ are, for arbitrary times, distributed
independently according to Poissonian distributions:
\begin{equation}
  \label{eq:p_s_u_pois}
  p_t(s, u)= e^{-K \alpha} \frac{(K \alpha_s(t))^{s}(K 
    \alpha_u(t))^{u}}{s! u!} 
\end{equation}
Again, this assumption is valid for $t=0$, whereas deviations appear
for larger times.  On average each variable is contained in $K
\alpha_s(t) = K(\alpha - \alpha_u(t))$ satisfied and $K \alpha_u(t)$
unsatisfied clauses.  If we plug this ansatz into
\eqref{eq:p_flip_walk} we get for an algorithm without greedy-steps
\begin{equation}
  \label{eq:p_u_s_u_pois}
  p^{(flip)}_t(s, u)= e^{-K \alpha} \frac{(K \alpha_s(t))^{s}(K 
    \alpha_u(t))^{u - 1}}{s! (u-1)!} 
\end{equation}
which again is a product of Poissonian distributions of $s$ and $u -
1$. Hence, on average, the negated variable $v^*$ is contained in $K
\alpha_s(t)$ satisfied and $K \alpha_u(t) + 1$ unsatisfied clauses.

\subsubsection{Random $K$-XOR-SAT}
\label{sec:poisson-xorsat}

We continue by first considering the analytically simpler case of
$K$-XOR-SAT. There, by flipping $v^*$, all $s$ satisfied clauses containing
$v^*$ become unsatisfied, whereas all $u$ unsatisfied  ones
become satisfied. 
The {\it expected number of unsatisfied clauses} $N_t^{(u)}$ changes
during one step as
\begin{equation}
  \label{eq:meansteppois}
  \Delta N_t^{(u)} =  -(K\alpha_u(t) + 1) + K\alpha_s(t) 
  = K\alpha - 2 K \alpha_u(t) - 1\ .
\end{equation}
Concentrating on the {\it average dynamics}, which is followed with
probability approaching one in the thermodynamic limit $N\to\infty$,
we have $N_t^{(u)} = N \alpha_u(t)$. Measuring the time $t$ in MC
sweeps, every algorithmic step contributes a $\Delta t = 1/N$, and the
difference on the left-hand-side of Eq. (\ref{eq:meansteppois}) can be
replaced by a time derivative (if $N\gg 1$),
\begin{equation}
  \label{eq:typtrajxor}
 \dot \alpha_u(t) = K\alpha - 2 K \alpha_u(t) - 1\ .
\end{equation}
If we solve this differential equation we get for the the energy
density of $K$-XOR-SAT
\begin{equation}
  \label{eq:energydensxorsatpoisc}
  \alpha_u(t)=\frac 1 {2 K} (K\alpha - 1 + C e^{-2 K t})
\end{equation}
In the typical starting configuration half the clause are satisfied
and half are not, i.\,e. $\alpha_u(0) = \alpha / 2$. So we finally get
\begin{equation}
  \label{eq:energydensxorsatpois}
  \alpha_u(t)=\frac 1 {2 K} (K\alpha - 1 +  e^{-2 K t})
\end{equation}
In Fig. \ref{fig:numtyp3xorsat} the results for different $\alpha$ are
compared to numerical simulations. For small times both curves
coincide, because correlations have not yet built up. Later the
algorithm reaches a lower density of unsatisfied clauses than the Poissonian
approximation would suggest.

We also see that there are two different regimes. For small $\alpha$
the energy decreases quickly to zero - reaching zero at finite MC
times with non-zero slope. For larger $\alpha$, the number of
unsatisfied clauses first decreases, but then reaches a positive
plateau value. Both regimes are separated by a dynamical threshold
which is located at
\begin{equation}
  \label{eq:dyntresholdxor}
  \alpha_{d} = \frac 1 K\ .
\end{equation}
In the special case $K=3$ we thus find $\alpha_d=1/3$ which coincides
perfectly with our numerical findings. Note that for
$\alpha<\alpha_d$, the algorithm thus constructs a satisfying
assignment already after a linear number of algorithmic steps. Above
$\alpha_d$, the algorithm does not reach a solution in linear times
with a probability tending to one in the large-$N$ limit.

\subsubsection{Random $K$-SAT}
\label{sec:poisson-sat}

For random $K$-SAT we can get a similar estimate. We have to take into
account that now satisfied clauses do not necessarily become
unsatisfied if a contained variable is inverted. For each $K$-SAT
clause $C$ there is one unsatisfying and $2^K - 1$ possible
satisfying assignments. The only case where the clause becomes
unsatisfied by flipping a single variable $v^*$ is the assignment
where this variable is the only correct assigned variable in $C$.  If
we assume independent clauses this happens with probability $1/(2^K -
1)$, so we get for the expected number of unsatisfied clauses
\begin{equation}
  \label{eq:meansteppoissat}
  \Delta N_t^{(u)} =  -(K \alpha_u(t) + 1) + \frac{1}{2^K - 1} K
  \alpha_s(t) = \frac{K \alpha}{2^K-1} - \frac{2^K K}{2^K - 1}
  \alpha_u(t) - 1
\end{equation}
Going for $N\to\infty$ again to continuous-time quantities and
differential equations, we find
\begin{equation}
  \label{eq:diffenergydenssatpois}
  \dot\alpha_u(t) = \frac{K \alpha}{2^K-1} - \frac{2^K K}{2^K - 1}
  \alpha_u(t) - 1
\end{equation}
with solution (the initial condition is given by $\alpha_u(0)=\frac
\alpha{2^K}$)
\begin{equation}
  \label{eq:energydenssatpoisa}
  \alpha_u(t) = \frac{1}{2^K K}\left(K \alpha +[2^K-1] 
    \left[e^{-\frac{2^K}{2^K-1}t}-1\right]\right)\ ,
\end{equation}
cf. Fig. \ref{fig:numtyp3sat}. For random $K$-SAT we thus find the
Poissonian estimate
\begin{equation}
  \label{eq:dyntresholdxorsat}
  \alpha_d = \frac{2^K - 1}{K}
\end{equation}
for the onset of exponential solution times. In the special case $K =
3$ we get $\alpha_d = 7 / 3$ which is smaller than the numerical value
2.7.

\subsection{Rate equation for the walksat algorithm}
\label{sec:rate}

We have seen that already a simple Poissonian approximation is able to
qualitatively reproduce the behavior of walk-SAT for linear solution
times, at least for a pure walk-dynamics without greedy steps. There
were, however, some systematic quantitative deviations, in particular
for the case of random $K$-satisfiability.  It is thus necessary to go
beyond the simple Poissonian ansatz for $p_t(s,u)$, i.e. for the
time-dependent fraction of Boolean variables belonging to exactly $s$
satisfied and $u$ unsatisfied clauses. Our aim is to work only with
these quantities, i.e. we still have to keep the approximation that
the joint distribution for variables within one clause factorizes.
This approximation of independent neighboring variables was already
used in the beginning of this section, when $p^{(flip)}(s,u)$ was
derived, cf. Eqs. (\ref{eq:p_flip_walk}-\ref{eq:pflip_greedy_k_3}).

\subsubsection{Random $K$-XOR-SAT}
\label{sec:rate_k-xor-sat}

As above, we denote by $N_t\left(s, u\right) = N p_t(s, u)$ the
expected number of variables that occur in exactly $s$ satisfied and
$u$ unsatisfied clauses at step $t$. Our algorithm starts at $t = 0$
and each step counts as $\Delta t$. We follow the procedure in
\cite{We} to describe the typical evolution of the algorithm.

A variable $v^*$ with $s^*$ satisfied and $u^*$ unsatisfied clauses is
flipped. This occurs with probability $p^{(flip)}_t(s^*, u^*)$. Three
different processes contribute to $N_{t+\Delta t}(s, u)$:
\begin{itemize}
\item \emph{Contribution by $v^*$:} The $s^*$ satisfied clauses become
  unsatisfied, whereas the $u^*$ unsatisfied clauses become satisfied.
  The number of variables characterized by $s^*$ satisfied, $u^*$
  unsatisfied clauses is thus decreased by one, the number of
  variables in $u^*$ satisfied and $s^*$ unsatisfied clauses is
  increased by one. This means that the expected number of variables
  $N_{t}(s^*,u^*)$ is decreased by $p^{(flip)}_t(s^*, u^*)$, and
  $N_t(u^*, s^*)$ is increased by the same amount.
\item \emph{Neighbors of $v^*$ in previously satisfied clauses}: The
  flipped variable $v^*$ occurs, on average, in $\la s\ra^{(flip)}_t$
  previously satisfied clauses, where $\la\cdot\ra^{(flip)}_t =
  \sum_{s,u}(\cdot) p^{(flip)}_t(s, u)$. Since each clause contains
  $K$ variables, and since random formulas are locally tree-like,
  there are on average $(K-1)\la s\ra^{(flip)}_t$ neighbors in
  previously satisfied clauses.
  
  All these clauses become unsatisfied. This means that for each other
  variable contained in these satisfied clauses, the number of
  satisfied clauses goes down by one, the number of unsatisfied
  clauses is increased by one. Taking into account that, according to
  the assumption of independent neighbors, these belong to $s$
  satisfied and $u$ unsatisfied clauses with probability $s
  p_t(s,u)/\la s \ra_t$, we conclude that $N_t(s,u)$ is, on average,
  decreased by $(K-1) \la s\ra^{(flip)}_t \frac{s p(s,u)}{\la s\ra_t
    }$. One out of these $s$ satisfied clauses is the one with the
  flipped variable $v^*$, so the decrease of $N_t(s,u)$ is now added to
  $N_t(s-1,u+1)$.
\item \emph{Neighbors of $v^*$ in previously unsatisfied clauses}:
  Analogously to the discussion in the last item one gets
  contributions to $N_t(s,u)$ for variables $v$ which occur together
  with $v^*$ in unsatisfied clauses.
\end{itemize}

Combining these processes we get an evolution equation for the
expected numbers $N_t(s,u)$ of variables appearing in exactly $s$
satisfied and $u$ unsatisfied clauses at time $t$:

\begin{eqnarray}
  \label{eq:rate_n_xor}
    N_{t+\Delta t}(s,u) &=& N_{t}(s,u) - p^{(flip)}_t(s,
    u) + p^{(flip)}_t(u, s)\notag\\
    &&+(K-1)\la s\ra^{(flip)}_t \left(-\frac{s p_t(s,u)}{\la s\ra_t }
     +\frac{(s+1) p_t(s+1,u-1)}{\la s\ra_t }\right)\notag\\
    &&+(K-1)\la u\ra^{(flip)}_t \left(-\frac{u p_t(s,u)}{\la u\ra_t }
     +\frac{(u+1) p_t(s-1,u+1)}{\la u\ra_t }\right)
\end{eqnarray}

Setting again $\Delta t = 1/N$ and replacing differences by
derivatives in the thermodynamic limit, 
\begin{eqnarray}
  \label{eq:diff}
  N_{t+\Delta t}(s,u)-N_{t}(s,u) &=&
  N\left(p_{t+\Delta t}(s,u) -p_{t}(s,u)\right) \nonumber\\
  &=& \frac{p_{t+\Delta t}(s,u)-p_{t}(s,u)}{\Delta t} \nonumber\\
  &\to& \frac d{dt} p_{t}(s,u) 
\end{eqnarray} 
we get a set of differential equations for $p_t(s,u)$: 

\begin{eqnarray}
  \label{eq:rate_ppunkt_xor}
  \dot p_{t}(s,u) &=& - p^{(flip)}_t(s,
    u) + p^{(flip)}_t(u, s)\notag\\
    &&+(K-1)\la s\ra^{(flip)}_t \left(-\frac{s p_t(s,u)}{\la s\ra_t }
     +\frac{(s+1) p_t(s+1,u-1)}{\la s\ra_t }\right)\notag\\
    &&+(K-1)\la u\ra^{(flip)}_t \left(-\frac{u p_t(s,u)}{\la u\ra_t }
     +\frac{(u+1) p_t(s-1,u+1)}{\la u\ra_t }\right)
\end{eqnarray}

In the typical initial configuration the probability of a clause to be
unsatisfied is $1/2$ and so $p_0(s, u)$ is given by equation
\eqref{eq:p_s_u_pois} with $\alpha_s(t) = \alpha_u(t) = 1/2$.
 
\begin{figure}[htbp]
  \begin{center}
    \includegraphics[clip, height=7cm]{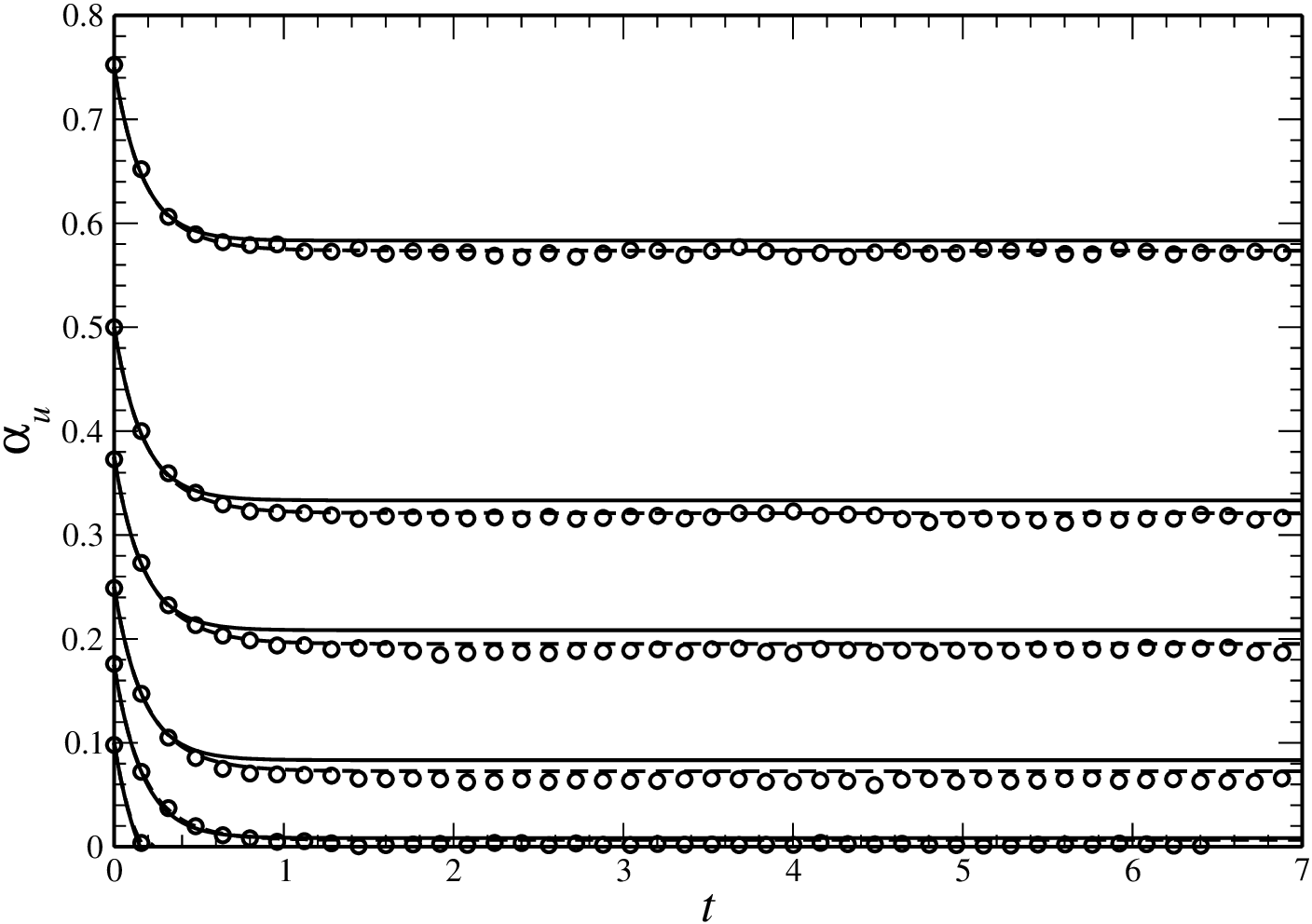}
    \caption{3-XOR-SAT: Typical number of unsatisfied clauses (divided
      by $N$), as a function of the MC time $t$, for walk-SAT
      with \emph{walk steps} only. Different
      ratios of $\alpha$ are shown, from top to bottom we have $\alpha
      = 1.5, 1, 0.75, 0.5, 0.35, 0.2$. The dashed line is obtained by
      numerically integrating equations \eqref{eq:rate_ppunkt_xor},
      the full line gives the Poissonian approximation. These results
      are compared to the evolution for a (random) single 3-XOR-SAT 
      instance with $N = 50000$, as given by the symbols.}
    \label{fig:numtyp3xorsat}
  \end{center}
\end{figure}

By numerical integration of \eqref{eq:rate_ppunkt_xor} we can find the
typical trajectory for an algorithm with given $p^{(flip)}_t$.  The
results for an algorithms without greedy-steps (e.g.  $p^{(flip)}_t(s,
u) = p^{(u)}_t(s,u)$) for different values of the ratio $\alpha=M/N$
are shown in figure \ref{fig:numtyp3xorsat}. They are compared with
numerical data obtained from single runs of the algorithm on a large
single, randomly selected sample formula. As we can see the
assumption of independent variables is suitable to describe the
behavior of the algorithm in this model. We also see that the
dynamical threshold $\alpha_d$, which marks the onset of exponential
solution times, is again given by 1/3.

\begin{figure}[htbp]
  \begin{center}
    \includegraphics[clip, height=7cm]{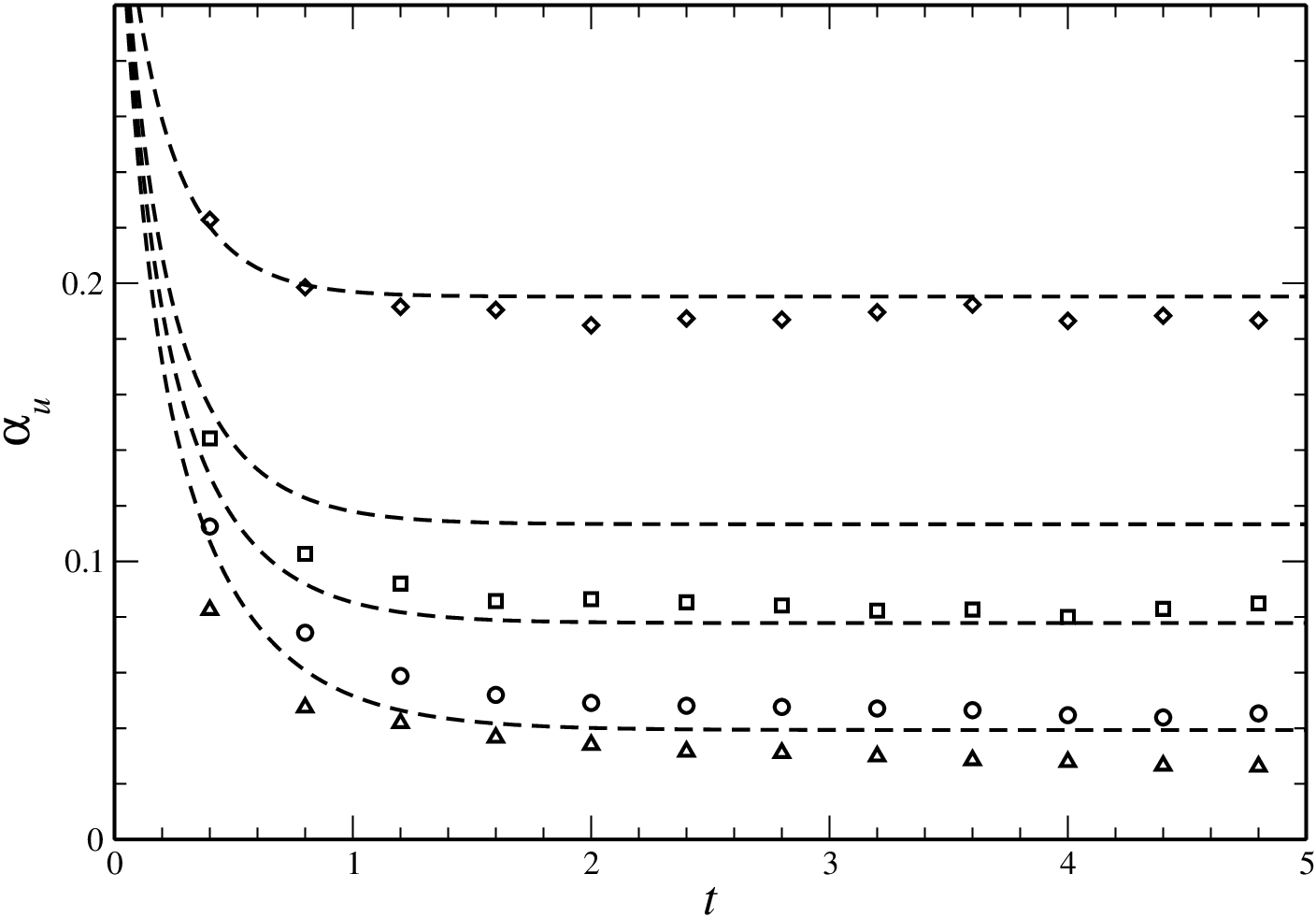}
    \caption{3-XOR-SAT:Influence of greedy steps on the behavior of
      the energy density at $\alpha = 0.75$. As above the dashed line
      is obtained by numerically integrating
      equation \eqref{eq:rate_ppunkt_xor} after plugging in
      eq. \eqref{eq:pflip_greedy_k_3} and using $N = 1/{\Delta
        t} = 50000$. The dotted line shows the evolution of a
      (random) single run of the
      algorithm with $N = 50000$. From top to bottom we have $q=0$
      (i.\,e. no greedy steps), $q=0.5$, $q=0.7$, $q=0.9$. The energy
      plateau decreases with q but due to correlations the integrated
      equation does not fit the numerical data.}
    \label{fig:xorgreedy}
  \end{center}
\end{figure}

When analyzing the algorithm including a fraction of greedy steps we
see that the assumption of independent variables is indeed very
crucial. In figure \ref{fig:xorgreedy} we show the result of the
numerical integration now using $p^{(flip)}(s,u)= q
p^{(flip-walk)}(s,u) + (1-q) p^{(flip-3-greedy)}_t(s, u) $ as given by
eq.  \eqref{eq:pflip_greedy_k_3}. Since in this case the probability
of a variable to be flipped depends on its neighbors naturally
correlations between neighboring variables built up. This explains why
the ansatz does not give a good quantitative approximation when greedy
steps are included.

\subsubsection{Random K-SAT}
\label{sec:rate_k-sat}

Let us now consider the slightly more involved case of random $K$-SAT.
As already discussed in the context of the Poissonian approximation,
we have to take into account that flipping a variable does not
necessarily unsatisfy all previously satisfied clauses the variable is
contained in. We assume again that the probability of such a clause to
become unsatisfied is clause- and time-independently given by its
naive average $\mu=1/(2^K-1)$. Similar to XOR-SAT we get three
contributions to $N_{t+\Delta t}(s, u)$, one coming from the flipped
variable itself, two from neighbors in previously satisfied (resp.
unsatisfied) clauses.
\begin{itemize}
\item If the flipped variable $v^*$ appears in exactly $s^*$ satisfied
  and $u^*$ unsatisfied clauses than, as in XOR-SAT, $N_t(s^*,u^*)$ is
  decreased by one. This happens with probability
  $p^{(flip)}_t(s^*,u^*)$.
  
  By flipping $v^*$, all $u^*$ previously unsatisfied clauses become
  satisfied. Out of the $s^*$ previously satisfied clauses, a random
  number $k$ remains satisfied, $s^*-k$ become unsatisfied, i.e.
  $N_t(u^*+k, s^*-k)$ is increased by one. There are ${s^* \choose
    k}$ possibilities for selecting these $k$ clauses, each one
  appearing with probability $\mu^{s^*-k}(1-\mu)^k$.

  The total contribution by $v^*$ is obtained by summing over all
  possible values of $k$.
\item The contributions from neighbors of the flipped variable are
  similar to XOR-SAT. The only difference is that the average number
  of neighboring variables on satisfied clauses becoming unsatisfied
  is now  $ \mu(K-1)\cdot{\la s\ra^{(flip)}_t }$. 
\end{itemize}

\begin{figure}[htbp]
  \begin{center}
    \includegraphics[clip, height=7cm]{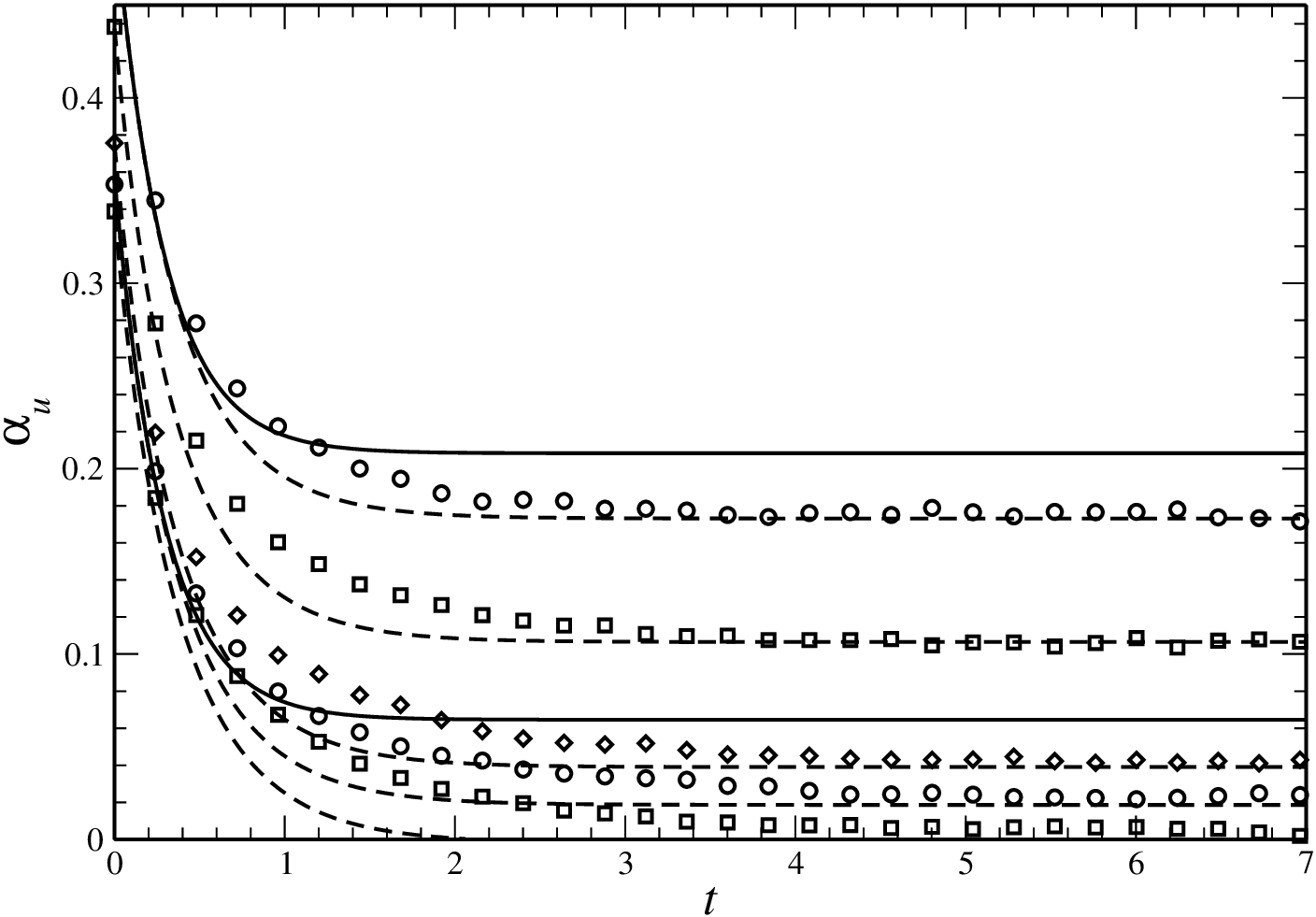}
    \caption{3-SAT: Running time of the Walksat-Algorithm  with
      \emph{walksteps} only. Different ratios of $\alpha$ are shown,
      from top to bottom we have $\alpha = 4.0, 3.5, 3.0, 2.85, 2.7$.
      The dashed line is obtained by integrating equations
      \eqref{eq:rate_ppunkt_sat} with $N = 1/{\Delta t} = 50000$. The
      symbols show the evolution of a (random) single run of the
      algorithm with $N = 50000$. The solid line shows the analytical
      solution \eqref{eq:energydensxorsatpois} of the Markov equation
      assuming a Poissonian distribution $p_t(s,u)$ for all times
      $t$, for clarity only $\alpha=4.0$ and $2.85$ are depicted.}
    \label{fig:numtyp3sat}
  \end{center}
\end{figure}

\begin{figure}[htbp]
  \begin{center}
    \includegraphics[clip, height=5.5cm]{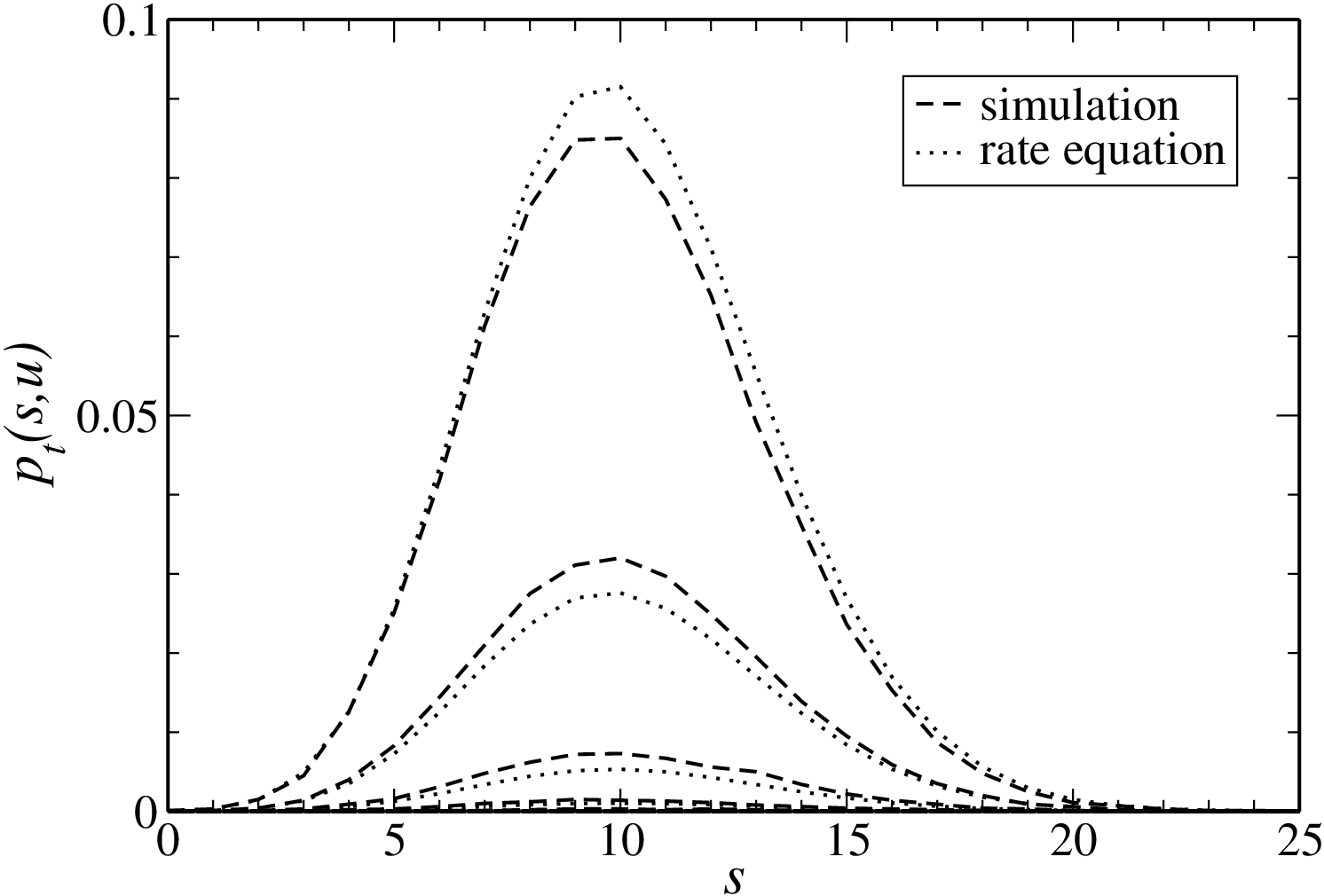}
    \includegraphics[clip, height=5.5cm]{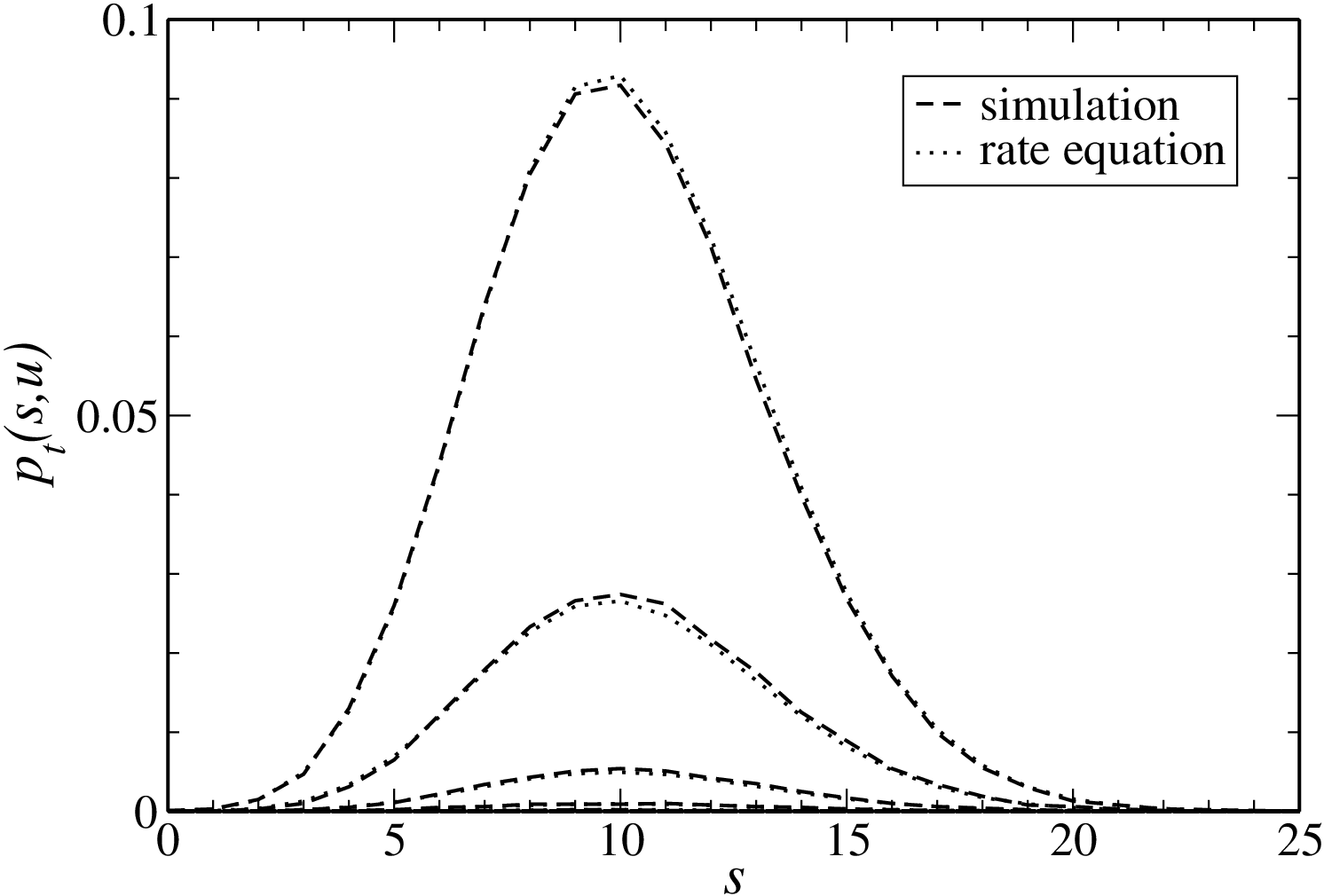}
    \caption{3-SAT: Distributions $p_t(s,u)$ for t=1.4 (left) and
      t=6. The results are shown as a function of $s$, the different
      curves correspond to $u=0,1,2,3$ (from top to bottom). One can
      see that numerical and analytical results differ for $t=1.4$,
      whereas they are very close for larger times corresponding to
      the energy plateau.}
    \label{fig:psudistr}
  \end{center}
\end{figure}

Combining all contributions we derive a set of differential equations
for the probability distribution of the variables:
\begin{eqnarray}
  \label{eq:rate_ppunkt_sat}
  \dot p_{t}(s,u) &=& - p^{(flip)}_t(s,u) +
  \left(\frac1{2^K-1}\right)^u \sum_{k=0}^{s} { u+k \choose k } \left(
    1 - \frac1{2^K-1}\right)^k p^{(flip)}_t(u+k,s-k) \notag\\
    &&+\frac{K-1}{2^K-1}\la s\ra^{(flip)}_t \left(-\frac{s p_t(s,u)}{
        \la s\ra_t }
     +\frac{(s+1) p_t(s+1,u-1)}{\la s\ra_t }\right)\notag\\
    &&+(K-1)\la u\ra^{(flip)}_t \left(-\frac{u p_t(s,u)}{\la u\ra_t }
     +\frac{(u+1) p_t(s-1,u+1)}{\la u\ra_t }\right)
\end{eqnarray}
Also these equations have to be solved numerically. The results for
the most interesting case $K=3$ (i.\,e. $q_u=1/(2^K - 1) = 1/7$) for
different values of $\alpha$ are shown in figure \ref{fig:numtyp3sat}.
Even if they are quantitatively much more accurate than the Poissonian
approximation, there are some systematic deviations compared to direct
numerical simulations. The curves match the simulation results for
small times. Then correlations between neighboring variables build up,
violating our basic assumption.  However, for larger times both curves
match again, because the same distribution $p_t(s,u)$ is reached. This
can be seen in the histogram in figure \ref{fig:psudistr}: At $t =
1.4$ the distributions $p_t(s,u)$ as derived from the rate equations
or evaluated numerically are different, while after $t = 6$ they again have
almost converged to the same distribution.

\begin{figure}[htbp]
  \begin{center}
    \includegraphics[clip, height=7cm]{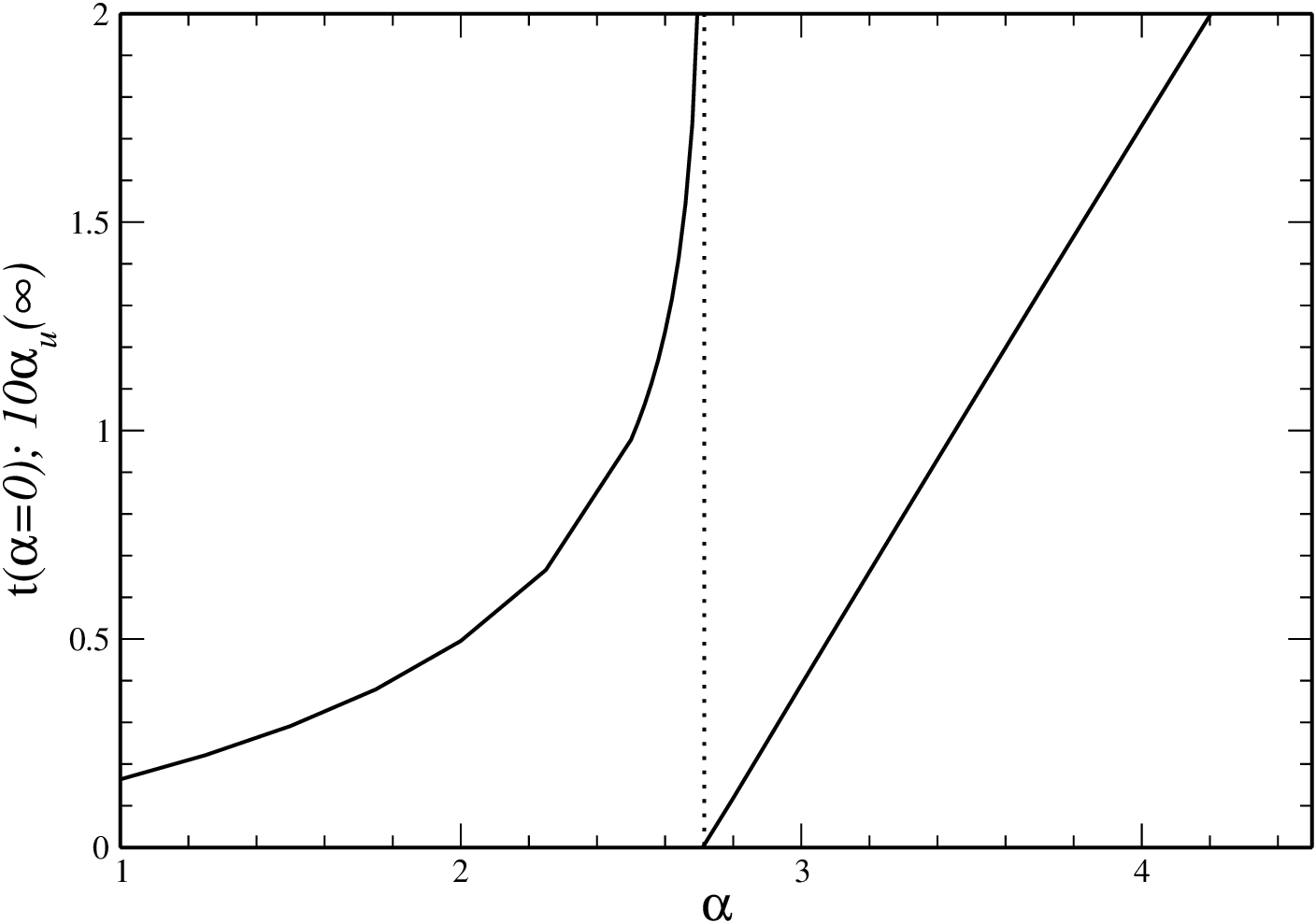}
    \caption{3-SAT: The left curve shows the (linear) solution time
      after which the expected energy density $\alpha_u(t)$ (from rate
      equations) reaches zero, as a function of $\alpha$. This time
      diverges logarithmically at $\alpha_d$. For larger $\alpha$, a
      non-zero energy plateau is found, which is shown in the right
      curve.}
    \label{fig:3sat_threshold}
  \end{center}
\end{figure}

This observation allows for a precise determination of the dynamical
threshold $\alpha_d$ which marks the transition from typically linear
to exponential algorithmic solution times needed by walk-SAT: The
transition is defined by the point where the expected energy density
$\alpha_u(t)$ asymptotically does not decrease to zero any more. In
figure \ref{fig:3sat_threshold} one can see that, for $K=3$, this
happens at $\alpha_d\simeq 2.71$.

\begin{figure}[htbp]
  \begin{center}
    \includegraphics[clip, height=7cm]{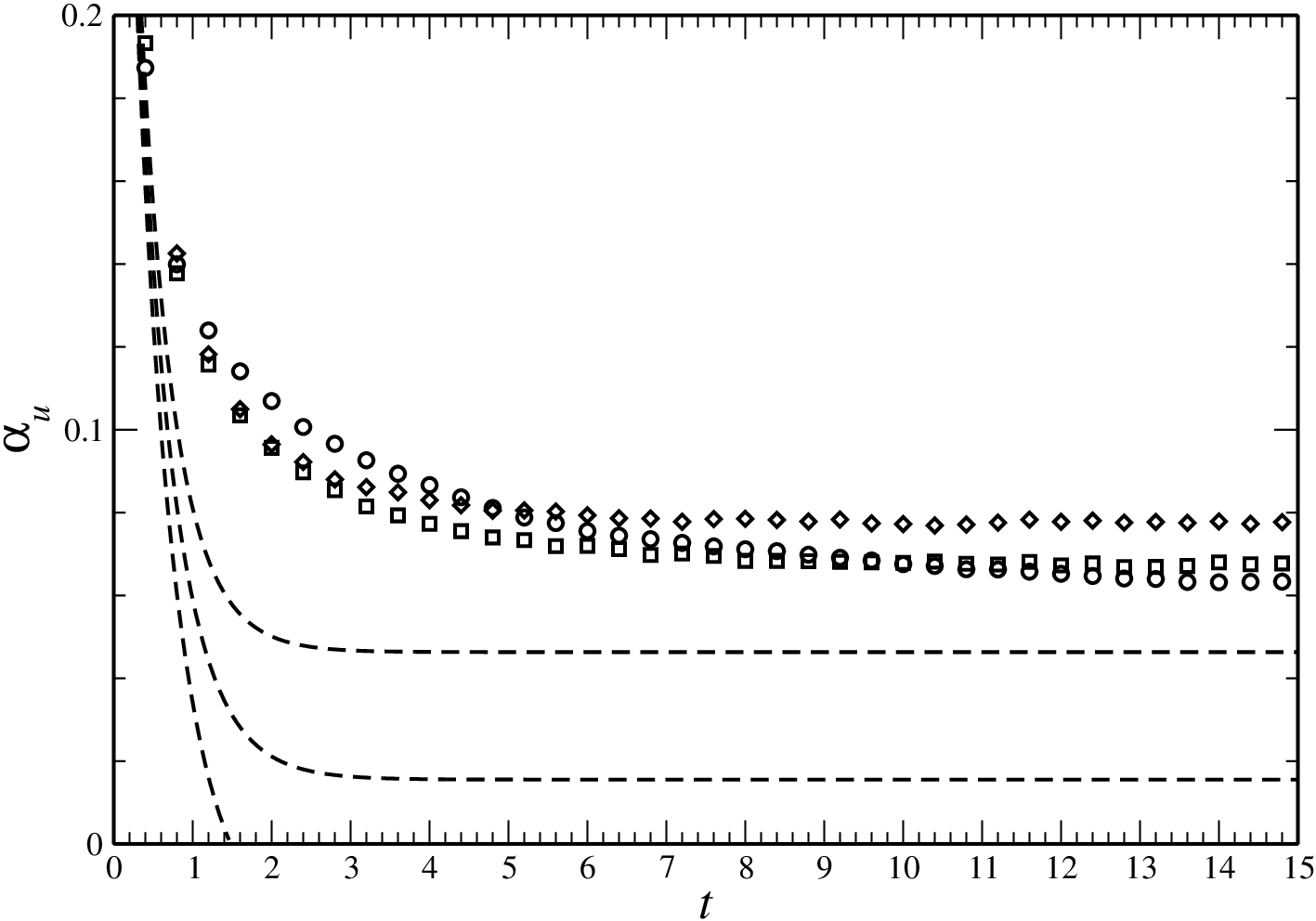}
    \caption{3-SAT: Influence of greedy steps  at $\alpha = 3.50$. As
      above the dashed line is obtained by numerically integrating
      equation \eqref{eq:rate_ppunkt_xor} after plugging in eq.
      \eqref{eq:pflip_greedy_k_3}. From top to bottom we have $q=0.5$,
      $q=0.7$, $q=0.9$.  The symbols show simulation data for the
      evolution of a single run of the algorithm with $N = 500000$. }
    \label{fig:3satgreedytime}
  \end{center}
\end{figure}

As already observed for XOR-SAT, the influence of greedy steps cannot
be reproduced very well. In figure \ref{fig:3satgreedytime} we show
results for three different $q$ at $\alpha = 3.5$. The energy density
obtained by assuming independent variables gives a too low energy
density. For $\alpha = 0.9$ it even decreases to zero at finite times,
contrary to our numerical results (cf. Sec. \ref{sec:num-ksat}). We
therefore conclude that the independent-neighbor approximation is only
suitable for the case without greedy steps, where less correlations
can be built up.

\section{Large deviations and the exponential-time behavior}
\label{sec:exp}

In the last section, we have characterized the \emph{typical
  linear-time behavior} of walk-SAT on satisfiable, randomly generated
$K$-SAT and $K$-XOR-SAT formulas. We have, within some approximation
assuming independent neighbors, calculated the trajectory which is
followed by the system in terms of the probabilities $p_t(s,u)$ that a
randomly selected variable belongs to exactly $s$ satisfiable and $u$
unsatisfiable clauses. ``Typical'' behavior means in this context that
the trajectory is followed with probability tending to one in the
thermodynamic limit $N\to\infty$. 

We have seen that there exists some (model-dependent) dynamical
threshold $\alpha_d$, below which the algorithm reaches zero energy,
i.e. a solution of the SAT formula, after linear time. Above
$\alpha_d$, the typical trajectory, however, shows a fast
equilibration towards a non-zero plateau value
$\alpha_u(t\to\infty)$. The walk-SAT algorithm is no longer able to
construct a solution in linear time, i.e. we expect the solution times
to become exponentially large. The approach of Sec. \ref{sec:lin} thus
fails to describe the final descent of the energy to zero.

In Sec. \ref{sec:num} we have seen that, for smaller system sizes, the
number of unsatisfied clauses fluctuates around its expected value.
Eventually these fluctuations become large enough that the system by
chance hits a solution -- fluctuations are the way  walk-SAT
finally succeeds constructing a solution. However, we expect these
fluctuations to be exponentially rare, i.e. we have to wait almost
surely an exponentially long time to really touch a solution.

This section is dedicated to characterizing these fluctuations, or, more
precisely, to calculating the probability
$\mathbb{P}(\alpha_u(0)\to\alpha_u(t_f)=0)$ that the system reaches
$\alpha_u(t_f)=0$ after some finite time $t_f$, under the condition
that the system started initially with some $\alpha_u(0)$. This
probability gives all important information about the dominant
exponential contribution to the typical running times $t_sol \simeq
e^{N\tau}$ beyond $\alpha_d$:
\begin{itemize}
\item For \emph{walk-SAT without restarts}, we start from a typical
  initial condition $\alpha_u(0)=1/2^K$ (for $K$-SAT) resp. 1/2 (for
  $K$-XOR-SAT), and we wait until the system reaches
  $\alpha_u(t)=0$. This does not happen for finite times, i.e. the
  solution time is given by the exponent
  \begin{equation}
    \label{eq:t_norestart}
    \tau \simeq - \lim_{t_f\to\infty} \lim_{N\to\infty} \frac 1N \ln
    {\mathbb{P}(\alpha_u(0)\to\alpha_u(t_f)=0)} 
  \end{equation}
  The solution time is thus, in general, exponentially large in $N$.
  Note that the order of limits in the above expression is relevant,
  there $t_f$ measures only a finite MC time scale. With interchanged
  limits, the right-hand side would vanish, since the algorithm finds
  a solution after exponential times $t_f$ with probability one.
\item For \emph{walk-SAT with restarts}, the situation changes
  slightly. Let us assume that the algorithm stops every $t_f N$
  walk-SAT iterations and re-initializes the variables randomly. In
  this case, we have to take into account two distinct rare events:
  First, the starting point may be close to a solution, i.e.
  $\alpha_u(0)$ is atypically small. This happens with probability
  $\rho(\alpha_u(o)) \sim e^{Ns(\alpha_u(0))}$ where $s(\alpha_u(0))$
  is the micro-canonical entropy for the energy density $\alpha_u(0)$.
  $\rho(\alpha_u(0))$ tends to one for the typical starting point
  discussed in the previous item, and becomes exponentially rare for
  smaller initial energies. This may be balanced by the fact, that
  finding a solution after some given time $t_f$ becomes more probable
  for smaller initial energies. From the probability of finding a
  solution after a single restart, $\rho(\alpha_u(0))\max_{0\leq t
    \leq t_f} {\mathbb{P}(\alpha_u(0)\to\alpha_u(t)=0)}$, we can read
  off the number of restarts $t_sol = e^{N\tau}$ needed to find a
  solution with high probability:
  \begin{equation}
    \label{eq:t_restart}
    \tau \simeq - \max_{\alpha_u(0)} \lim_{N\to\infty} \frac 1N \ln \left[
        \rho(\alpha_u(0)) \max_{0\leq t \leq t_f}
        \mathbb{P}(\alpha_u(0)\to\alpha_u(t)=0) 
      \right]
  \end{equation}
\end{itemize}
Our aim is thus to calculate the large-deviation functional
determining $\mathbb{P}(\alpha_u(0)\to\alpha_u(t_f)=0)$. As we will
see, this can be done only within the Poissonian approximation, i.e.
throughout this section we assume
\begin{equation}
  \label{eq:p_s_u_pois2}
  p_t(s, u)= e^{-K \alpha} \frac{(K \alpha_s(t))^{s}(K 
    \alpha_u(t))^{u}}{s! u!} 
\end{equation}
with $\alpha_u(t)+\alpha_s(t)=\alpha$ being time-independent.

\subsection{Random $K$-XOR-SAT}
\label{sec:exp_xorsat}

Here we discuss only an algorithm without greedy-steps, where the
above approximation works reasonably well. Therefore
$p^{(flip)}_t(s,u)$ is given by equation \eqref{eq:p_u_s_u_pois}. The
number of unsatisfied clauses in a formula at time $t$ is given by
$N\alpha_u(t)$. This number changes by $\Delta e = s-u$ in the next
step if a variable with $u$ unsatisfied and $s$ satisfied clauses is
flipped.  The probability $P(\Delta e)$ of a given energy shift
$\Delta e$ in a single step is consequently given by
\begin{eqnarray}
\label{ptdeltae}
  P_t(\Delta e) &=& \sum_{s,u=0}^{\infty} p^{(flip)}_t(s,u)
  \delta_{\Delta e, s-u} \notag\\
  &=& \sum_{s,u=0}^{\infty} \frac{u p_t(s,u)}{K\alpha_u(t)}
  \delta_{\Delta e, s-u} \notag\\
  &=& \sum_{s,u=0}^{\infty} e^{-K \alpha} 
  \frac{(K \alpha_s(t))^{s}(K \alpha_u(t))^{u}}{s! u!}
  \frac{u}{K\alpha_u(t)}\delta_{\Delta e, s-u} \notag\\
  &=& \sum_{s,u'=0}^{\infty} e^{-K \alpha} \frac{(K \alpha_s(t))^{s}(K 
    \alpha_u(t))^{u'}}{s! u'!}\delta_{\Delta e, s-u'-1}\ .
\end{eqnarray}

The probability $P_{\Delta T,t}(\Delta E)$ of a change of the number
of unsatisfied clauses by $\Delta E$ after $\Delta T$ steps is given
by the convolution of the single-step probabilities. For $\Delta T =
N\delta t$ with small $\delta t \ll 1$, the energy density
$\alpha_u(t)$ and thus $P_t(\Delta e)$ are almost time-independent, so
we get in Fourier space
\begin{eqnarray}
  \hat P_{\Delta T,t}(l) &=& \left(\hat P_t(l)\right)^{\Delta T} 
  = \left(\sum_{s,u=0}^{\infty} e^{-K \alpha} \frac{(K
    \alpha_s(t))^{s}(K \alpha_u(t))^{u}}{s! u!}\exp\left\{-i
    l(s-u-1)\right\}\right)^{\Delta T}\notag\\
  &=& \left(\exp\left\{-K\alpha + K\alpha_s(t) e^{-i l} + K\alpha_u(t)
        e^{i l}  + i l
      \right\}\right)^{\Delta T}
\end{eqnarray}
Switching again to intensive quantities, we have $\Delta E = N
\dau\delta t$ and thus
\begin{eqnarray}
\label{eq:problagrangian}
P_{\Delta T,t}(\Delta E) &=&  \int\frac{dl}{2\pi} e^{i l \Delta E} 
  \left(\hat P_t(l)\right)^{\Delta T} \notag\\
&=& \int\frac{dl}{2\pi} \exp\left\{N\delta t\left( i l\dau  -K\alpha 
  + K(\alpha-\alpha_u(t)) e^{-i l} + K\alpha_u(t)
        e^{i l}  + i l  \right)\right\}
\end{eqnarray}
as the probability of the algorithm for getting from energy density
$\alpha_u = E / N$ at time $t$ to energy density $(E - \Delta E) / N$
at time $t+\Delta T$.  To calculate the transition probability between
$\alpha_u(0)$ and arbitrary $\alpha_u(t_f)$ after linear time $t_f N$
we write $t_f$ as a composition of many small intervals $\delta t$. We
then get that transition probability by integrating over all possible
paths $\alpha_u(t)$ going from $\alpha_u(0)$ to $\alpha_u(t_f)$. By
this step also the conjugate variable $l$ becomes a time-dependent 
function $l(t)$,
\begin{equation}
  \label{eq:pathintxorpois}
  \mathbb{P}(\alpha_u(0)\rightarrow\alpha_u(t_f)) = 
  \int_{\alpha_u(0)}^{\alpha_u(t_f)} {\cal  D}\alpha_u(t)\int{\cal
    D}l(t)  \exp\left\{- N\int_0^t\delta t 
    {\cal L}(l(t),\au,\dau)\right\}\ ,
\end{equation}
where the Lagrangian $\cal L$ is given by
\begin{equation}
  \label{eq:lagrangianxor}
  {\cal L}(l(t),\au,\dau) = - i\kDE\dau  +K\alpha - K(\alpha-\au) 
   e^{-i\kDE} - K\alpha_u(t) e^{i\kDE}  - i\kDE 
\end{equation}
We can replace the integral by its saddle point in the thermodynamic
limit. Since $l(t)$ is not a dynamic variable ($\dot l(t)$ does not 
appear in the Lagrangian) we find
\begin{equation}
  \label{eq:dLdl}
0  = \frac{\partial {\cal L}}{\partial l} = i \dau - i K(\alpha-\au)
e^{-i\kDE} +  i K\alpha_u(t) e^{i\kDE}  + i \ .
\end{equation}
The saddle point in $\alpha_u(t)$ is given by the Euler-Lagrange
equation
\begin{equation}
  \label{eq:dLdepsilon}
0 =  \frac d{dt} \frac{\partial {\cal L}}{\partial \dot\alpha_u} -
\frac{\partial {\cal L}}{\partial \alpha_u} 
= i\dot l(t) + K e^{-i\kDE} - K e^{i\kDE} \ .
\end{equation}
We are, in particular, interested in trajectories leading to a
solution of the formula, i.\,e.\ trajectories starting at some
$\alpha_u(0)$ and going to $\alpha_u(t_f) = 0$ after some given final
time $t_f$. This results in a set of two coupled first-order
non-linear differential equations for $\alpha_u(t)$ and $\kDE$ with
two boundary conditions given for $\alpha_u(t)$, and none for $\kDE$.
By substituting $\kappa(t)=e^{i\kDE}$ the equations read
\begin{eqnarray}
  \label{eq:eqsetxorsat}
  \dot\alpha_u(t) &=& -1 -
  K\alpha_u(t)\kappa(t)+K\frac{\alpha-\alpha_u(t)}{\kappa(t)}\notag\\
\dot\kappa(t)&=&K \kappa^2(t) - K
\end{eqnarray}
A trivial solution of the second equation, $\kappa(t)\equiv 1$ leads
to $\dot\alpha_u(t) = -1 - K\alpha_u(t) + K (\alpha-\alpha_u(t))$
which is exactly the equation for the typical trajectory given by
\eqref{eq:typtrajxor}.  Indeed we have ${\cal L}(\kappa\equiv
1,\alpha_u,\dot\alpha_u) \equiv 0$, so this trajectory has probability
1 in the thermodynamic limit.

This solution is, however, not stable since we have
$\dot\kappa(t) < 0$ for $\kappa(0) < 1$ and $\dot\kappa(t) > 0$ for
$\kappa(0) > 1$, i.e. the
trajectory deviates from the typical one once $\kappa$ deviates from
1. We can, however, solve the equations for XOR-SAT in this
Poissonian approximation and get
\begin{eqnarray}
  \label{eq:solxorsatpois}
\kappa(t) &=& \frac{1+A e^{2 K t}}{1-A e^{2 K t}}\notag\\
\alpha_u(t) &=& \alpha_u(0) e^{-2 K t}
\frac{1-A^2 e^{4Kt}}{1-A^2}\notag\\
&&+\int_0^td\tau\left(-1+K\alpha\frac{1-Ae^{2Kt}}{1+Ae^{2Kt}}\right)
e^{-2K(t-\tau)}\frac{1-A^2e^{4Kt}}{1-A^2e^{4K\tau}}\ .
\end{eqnarray}
In principle also the integrals in the second expression can be
carried out analytically, but we failed to find a compact
representation of the result. The solution still contains the unknown
constant $A$ which has to be adjusted to meet the final condition 
$\alpha_u(t_f)=0$. We have observed that $A$ is slightly smaller
than $e^{-2Kt_f}$, but it is easier to determine $t_f(A)$ than its
inverse $A(t_f)$.

\begin{figure}[htbp]
  \begin{center}
    \includegraphics[clip, height=7cm]{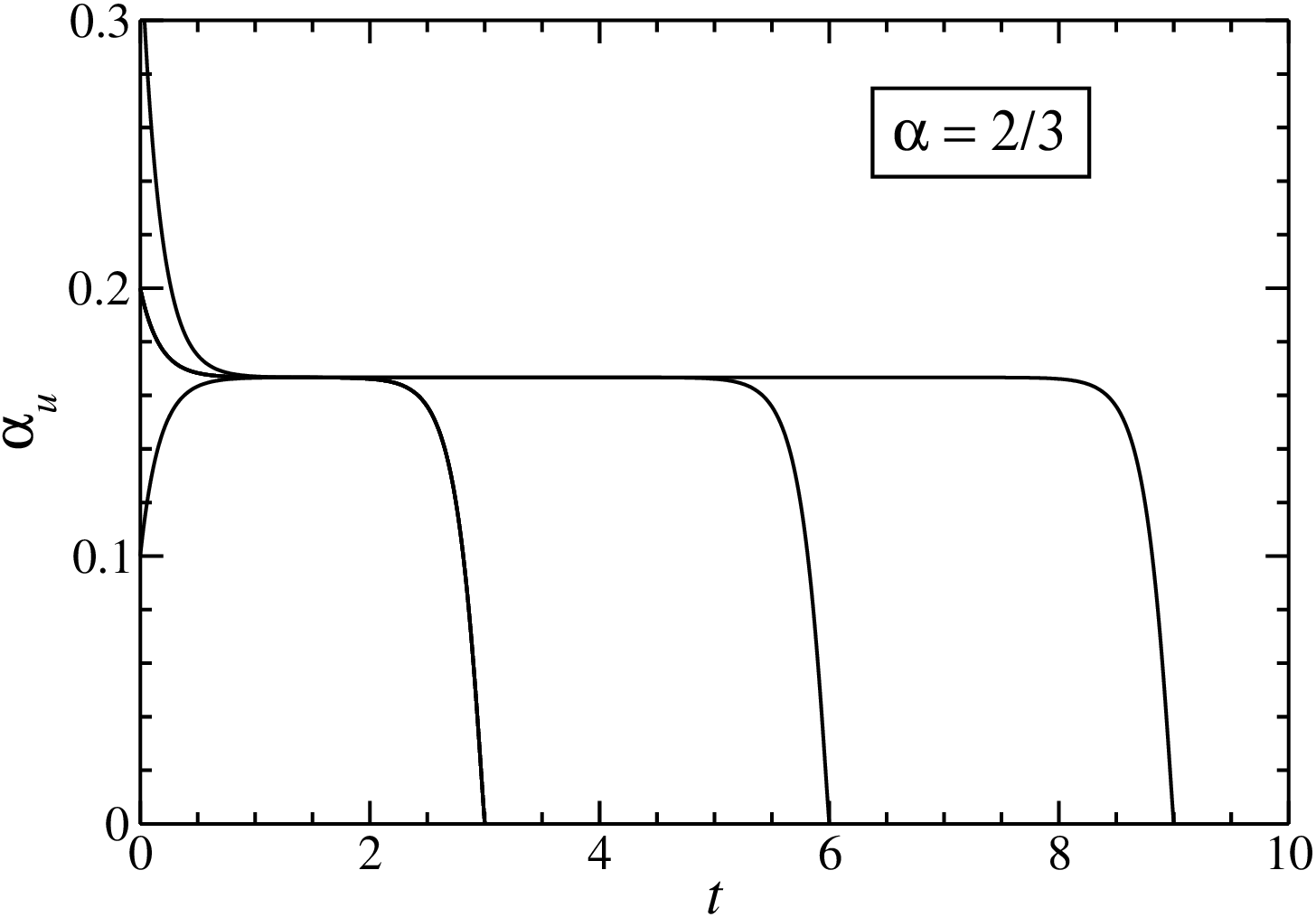}
    \caption{3-XOR-SAT at $\alpha=2/3$: Energy densities $\au$ for
      various initial conditions $\alpha_u(0)$ and solution times
      $t_f$. The system first equilibrates to a plateau being
      independent on the initial condition, and finally solves the 
      SAT formula by a macroscopic fluctuation.}
    \label{fig:3hsatendens}
  \end{center}
\end{figure}

The trajectories show an interesting behavior, cf. figure
\ref{fig:3hsatendens}: After about 1 MC-sweep the energy reaches a
plateau independent of the starting energy density $\alpha_u(t)$. The
plateau value is the same as given by the typical trajectory and
almost independent of the time $t_f$ where the solution is found. The
energy drops to $0$ suddenly about 1 MC-sweep before $t_f$. This is
similar to the qualitative picture we observed numerically in Sec.
\ref{sec:num}: The system first equilibrates and then, by means of an
exponentially improbable fluctuation, reaches zero energy, cf. fig.
\ref{fig:walksattimefluct}.  The fluctuations which are present in the
numerical data cannot be seen in the analytical curve due to the fact
that the latter one gives an average over all possible trajectories
under the condition that $\alpha_u=0$ is reached for the first time at
$t_f$, so only the very last fluctuation leading to the solution is
common to all possible numerical trajectories.

\begin{figure}[htbp]
  \begin{center}
    \includegraphics[clip, height=7cm]{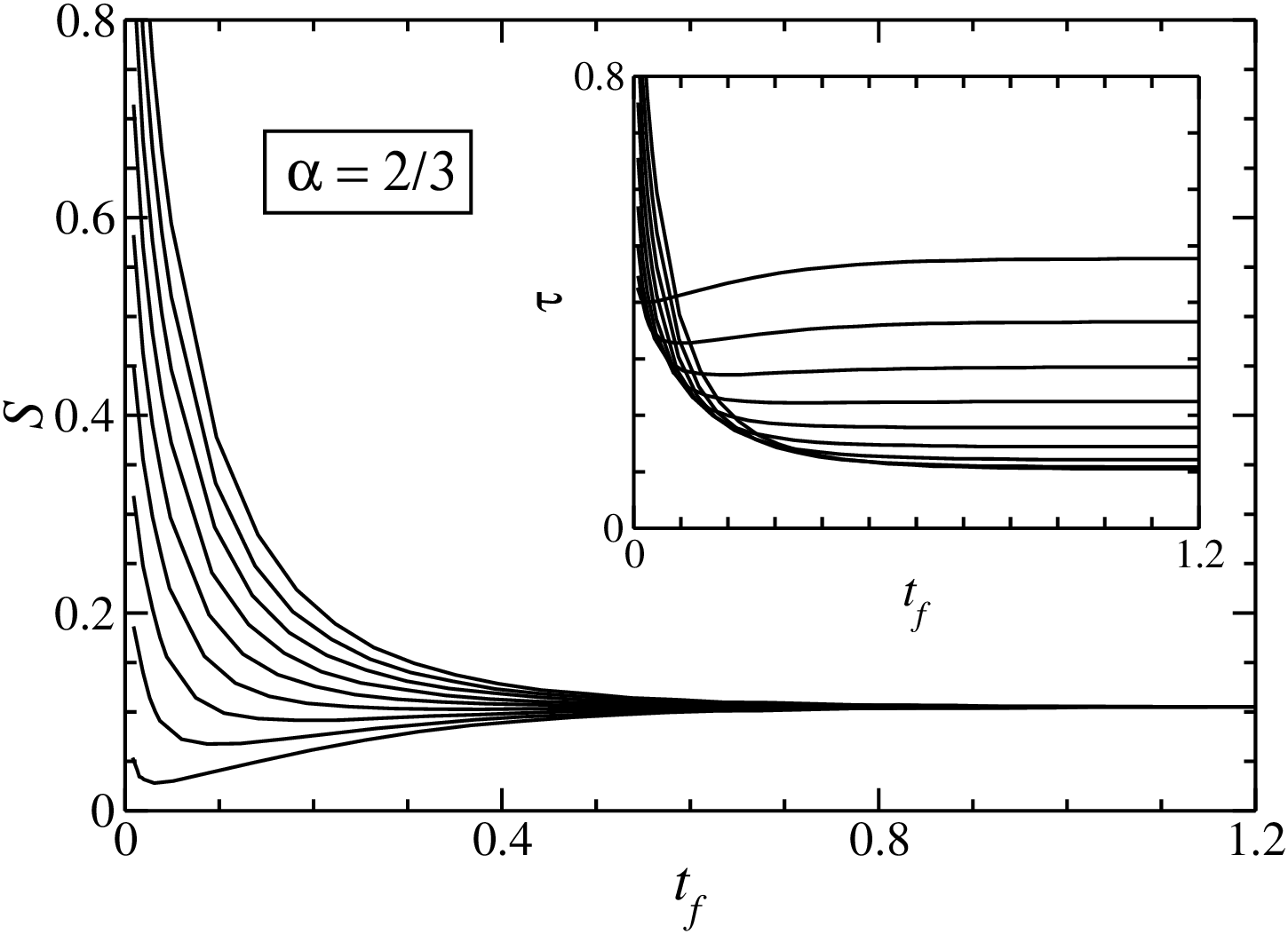}
    \caption{3-XOR-SAT at $\alpha=2/3$: 
      Action $S$ as a function of the resolution time $t_f$, for
      initial conditions $\alpha_u(0)=0.02,0.06,0.1,...,0.34$, from
      bottom to top. The inset shows the logarithm of the predicted
      solution time for the same values $\alpha_u(0)$, but now from
      top to bottom.}
    \label{fig:3hsatwirk}
  \end{center}
\end{figure}

To calculate the probability that the algorithm, starting at some
$\alpha_u(0)$, finds a solution after time $t_f$ we have to calculate
the action
\begin{eqnarray}
  \label{eq:actionxorsat}
S({\cal L}(\kappa(t),\alpha_u(t),\dot\alpha_u(t))) &=& \int_0^{t_f}
  dt {\cal L}(\kappa(t),\alpha_u(t),\dot\alpha_u(t))\nonumber\\
&=& \int_0^{t_f}
  dt\left( - \log(\kappa(t))(\dau +1)  +K\alpha -
    K\frac{\alpha-\au}{\kappa(t)} - K\alpha_u(t) \kappa(t)\right),
\end{eqnarray}
using solution (\ref{eq:solxorsatpois}). The evaluation is simplified
by plugging in the saddle-point equations in order to eliminate
$\dot\alpha_u(t)$,
\begin{eqnarray}
S({\cal L}(\kappa(t),\alpha_u(t),\dot\alpha_u(t))) &=& K \int_0^{t_f}
  dt\left( - \log(\kappa(t))
    (-\alpha_u(t)\kappa(t)+\frac{\alpha-\au}{\kappa(t)})
 +\alpha -
    \frac{\alpha-\au}{\kappa(t)} - \alpha_u(t)
    \kappa(t)\right)
\end{eqnarray}
The results are shown in Fig \ref{fig:3hsatwirk} for different values
of the initial condition $\alpha_u(0)$ and different solution times.
For the typical initial condition $\alpha_u(0)=\alpha/2$ we find a
monotonically decreasing function which has practically reached its
asymptotic value for $t_f>1$. From equation \eqref{eq:pathintxorpois},
the probability that the algorithm finds a solution is given by
\begin{equation}
  \label{eq:probaction}
  \mathbb{P}(\alpha_u(0)\rightarrow\alpha_u(t_f) = 0)
  = \exp\left\{- N S\right\},
\end{equation}
and the typical solution time of the algorithm \emph{without restarts} 
is given by Eq. (\ref{eq:t_norestart}),
\begin{equation}
  \label{eq:soltime}
  t_{sol} = \lim_{t_f\to\infty} e^{N S}\ .
\end{equation}

We also observe that, for smaller than typical $\alpha_u(0)$, the
action shows a pronounced minimum for small solution times. This
minimum corresponds to trajectories which start close to a solution
(small $\alpha_u(0)$) and go more or less directly to this solution
(small $t_f$). As discussed in the beginning of this section, it may
be possible that the algorithm can profit from this by using
random restarts. Taking the entropy as calculated in Ref.
\cite{RiWeZe}, we however find that the minimum in $S$ is
over-compensated by the small entropy of low-energy starting
configurations, cf. the inset of Fig.  \ref{fig:3hsatwirk} where
$\tau(\alpha_u(0),t_f) = -1/N\ \ln( \rho(\alpha_u(0)) \mathbb{P} )$ is
presented.  The minimum of $\tau$ is still found for the typical
starting configuration, and it is related to the typical running time
by $t_{sol}=\min e^{N\tau(\alpha_u(0),t_f)}$. Here it coincides with
the solution time without restarts.

\begin{figure}[htbp]
  \begin{center}
    \includegraphics[clip, height=7cm]{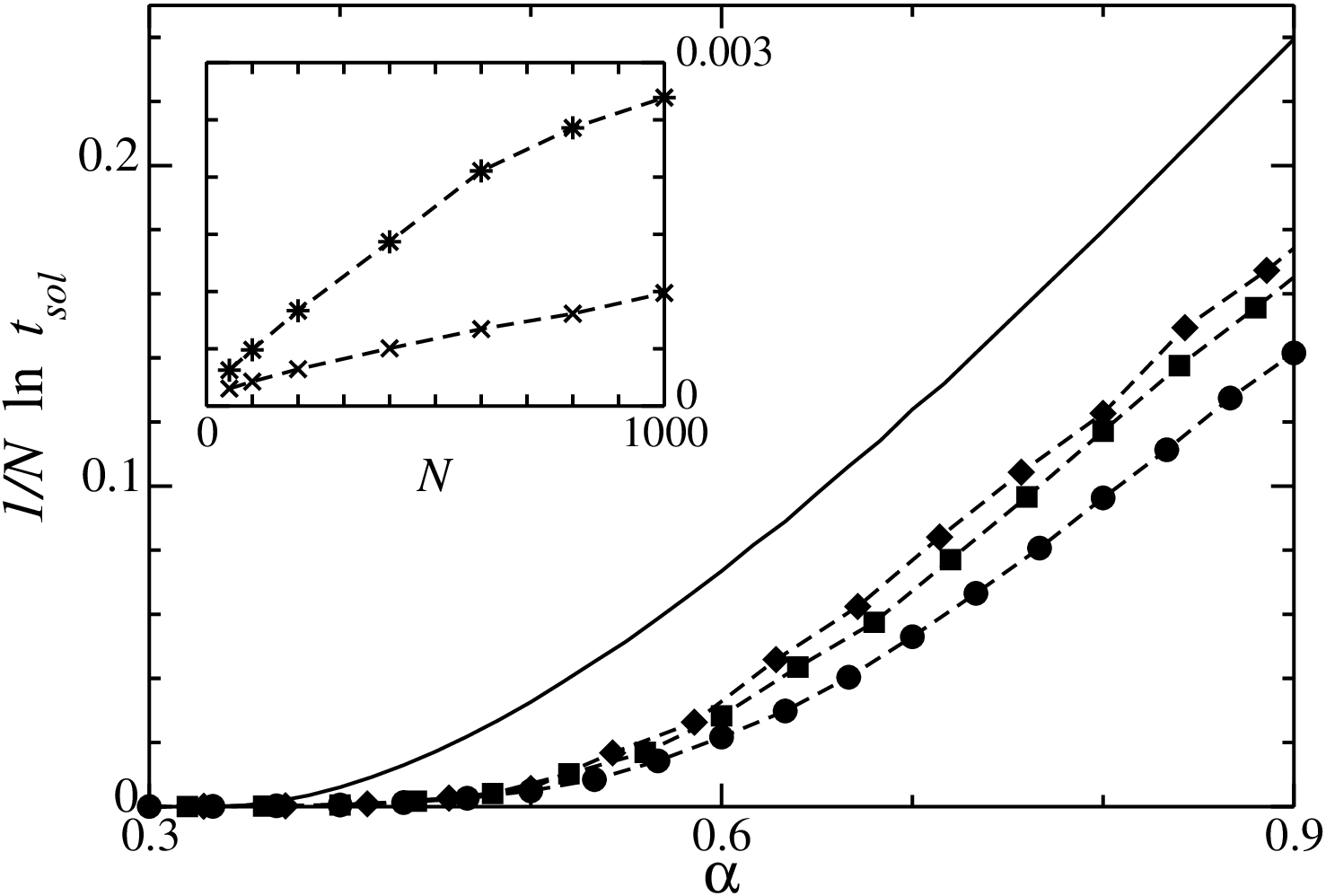}
    \caption{3-XOR-SAT: 
      Solution time $t_{sol}$ for Sch\"oning's algorithm (only walk
      steps, random restarts after $t_fN=3N$ steps) measured as the
      number of restarts, as a function of $\alpha$. The analytical
      result is given by the full line. Numerical data for
      $N=30,50,70$ (dots, squares, diamonds, lines are guides to the
      eyes only) seem to indicate much smaller solution times.  The
      inset shows, however, that there are huge finite size effects
      for $\alpha=0.4, 0.42$ (crosses, stars). The analytical
      estimates for the corresponding solution times are
      $\ln(t_{sol})/N \simeq 0.0061, 0.0099$.  }
    \label{fig:3hsattime}
  \end{center}
\end{figure}

In Fig. \ref{fig:3hsattime}, the resulting solution time is compared to
numerical data obtained using the algorithm with random restarts after 
$3N$ iterations. Due to the exponential behavior only small systems up
to $N=70$ could be investigated in the full satisfiable region. The
resulting running times seem to be much smaller than the analytical
predictions. There are, however, huge finite size effects. In the
inset we show numerical data for $\alpha=0.4$ and $0.42$, where the
exponent is still small enough that systems up to $N=1000$ can be
easily solved. It is obvious that even from such large systems the
asymptotic running time cannot be reasonably estimated. On the other
hand, the qualitative behavior is well-represented by the analytical
curve, in particular the sub-linear slope close to the threshold.
The analytical curve suggests $\ln t_{sol}\sim
\sqrt{\alpha-\alpha_d}$. Another interesting observation is that, at
the SAT/UNSAT threshold $\alpha_c=0.918$, the analytical prediction
for the solution-time exponent is 0.249, which is smaller but quite
close to Sch\"oning's rigorous upper worst-case bound
$\ln(4/3)\simeq0.288$.

\subsection{Random $K$-SAT}
\label{sec:exp_sat}

\begin{figure}[htbp]
  \begin{center}
    \includegraphics[clip, height=7cm]{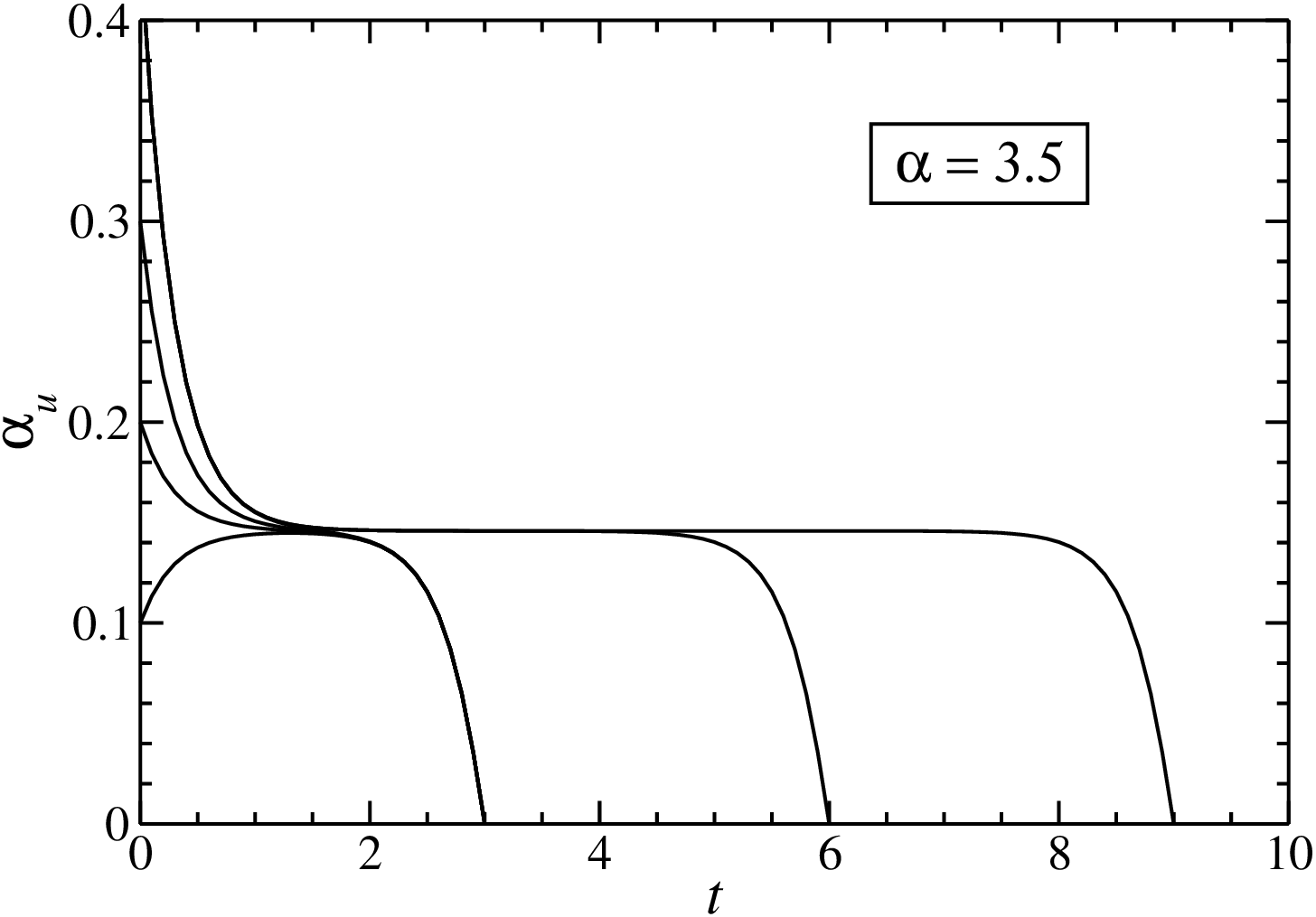}
    \caption{3-SAT at $\alpha=3.5$: Energy densities $\au$ for various
      initial conditions $\alpha_u(0)$ and solution times $t_f$. The
      system first equilibrates to a plateau being independent on
      the initial condition, and finally solves the SAT formula by a
      macroscopic fluctuation.}
    \label{fig:3satendens}
  \end{center}
\end{figure}

\begin{figure}[htbp]
  \begin{center}
    \includegraphics[clip, height=7cm]{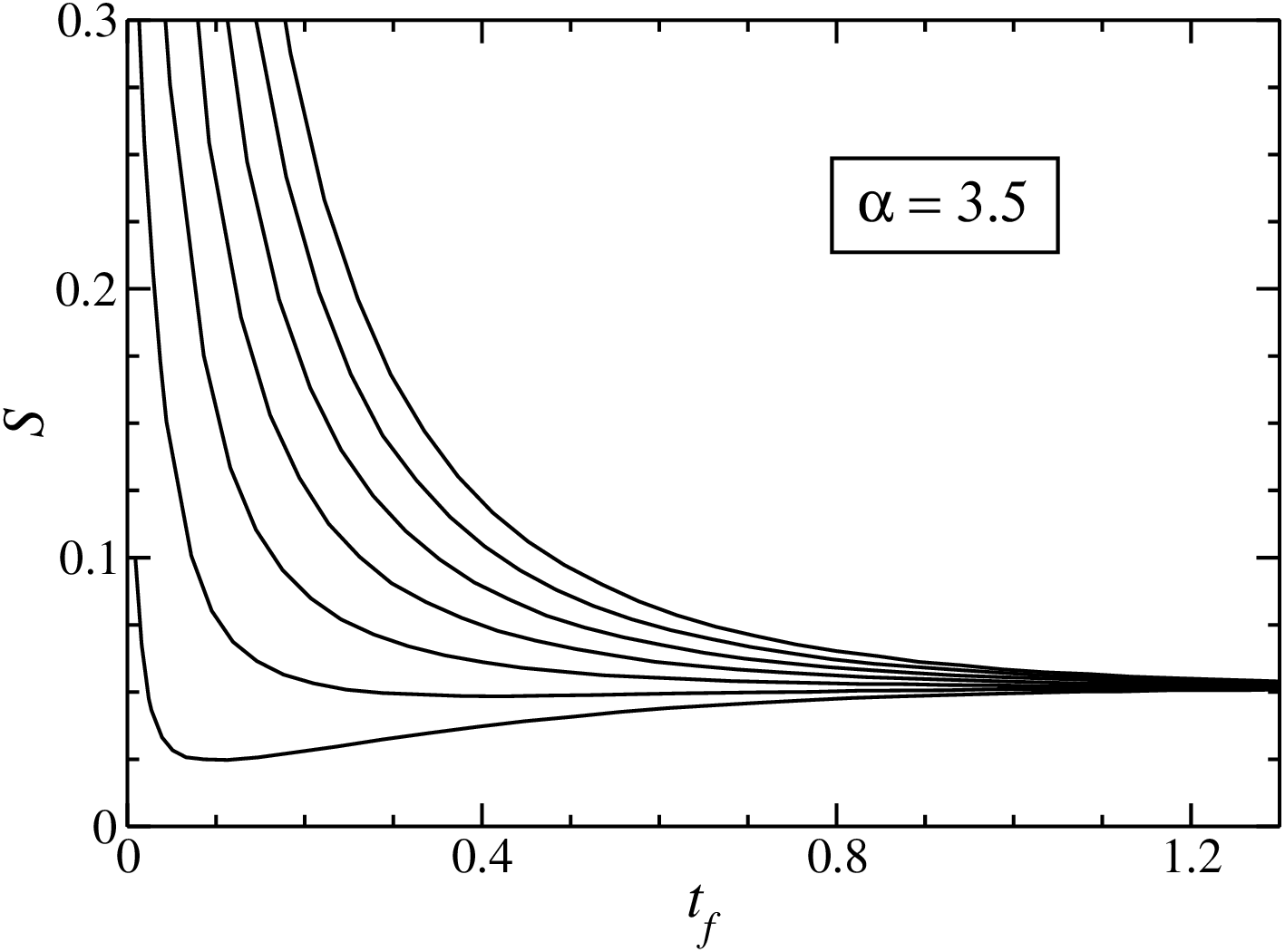}
    \caption{3-SAT at $\alpha=3.5$: 
      Action $S$ as a function of the resolution time $t_f$, for
      initial conditions $\alpha_u(0)/\alpha=0.1,0.3,0.5,...,0.13$, from
      bottom to top.}
    \label{fig:3satwirk}
  \end{center}
\end{figure}

The same type of analysis can be done for the case of random $K$-SAT.
The main difference is, as mentioned already in Sec. \ref{sec:lin},
that a satisfied clause does not necessarily become unsatisfied when
one of its variables is flipped. This happens only if the clause is
satisfied only by the variable to be flipped, which is one of the
$2^K-1$ satisfying assignments to this clause. We again use the
assumption that the variables in one clause are uncorrelated and
assume that a clause becomes unsatisfied with probability
$\mu=1/(2^K-1)$. In analogy to the discussion above we conclude that
the probability that a variable flip leads to a given energy change
$\Delta e$ is given by
\begin{equation}
  \label{eq:pqdeltaesat}
P_t(\Delta e) = \sum_{s,u=0}^{\infty} p^{(flip)}_t(s,u) \sum_{k=0}^s
 {s \choose k} \mu^k (1-\mu)^{s-k} \delta_{\Delta e, k-u} 
\end{equation}
where $k$ sums over all possible numbers of clauses which become
unsatisfied in the considered algorithmic step. Concentrating again on
the pure walk algorithm without greedy steps, i.e. on $q=1$, we can
go through the same procedure as for $K$-XOR-SAT. The transition
probability from some initial to some final density of unsatisfied
clauses is, in the Poissonian approximation (\ref{eq:p_s_u_pois2})
given by the path integral
\begin{equation}
  \label{eq:pathintpois}
  \mathbb{P}(\alpha_u(0)\rightarrow\alpha_u(t_f)) = 
  \int_{\alpha_u(0)}^{\alpha_u(t_f)} {\cal  D}\alpha_u(t)\int{\cal
    D}\kappa(t)  \exp\left\{- N\int_0^t\delta t 
    {\cal L}(\kappa(t),\au,\dau)\right\}\ ,
\end{equation}
with the modified Lagrangian
\begin{equation}
  \label{eq:lagrangian}
  {\cal L}(\kappa(t),\au,\dau) = - (1+\dau) \ln(\kappa(t)) +K\alpha 
   - K(\alpha-\au) \left(1-\mu +\frac \mu {\kappa(t)} \right) 
   - K\alpha_u(t) \kappa(t) \ . 
\end{equation}
The saddle-point equations are given by
\begin{eqnarray}
  \label{eq:eqsetxorsatsaddle}
  \dot\alpha_u(t) &=& -1 -
  K\alpha_u(t)\kappa(t)+K\mu\frac{\alpha-\alpha_u(t)}{\kappa(t)}\notag\\
\dot\kappa(t)&=&K \kappa^2(t) -K(1-\mu)\kappa(t) - K\mu\ .
\end{eqnarray}
Their solution dominates, for $N\to\infty$, the path integral
(\ref{eq:pathintpois}) and is given by the generalization of Eqs.
(\ref{eq:solxorsatpois}):
\begin{eqnarray}
  \label{eq:solsatpois}
\kappa(t) &=& \frac{1+ \mu  A e^{2 K t}}{1-A e^{2 K t}}\notag\\
\alpha_u(t) &=& \alpha_u(0) e^{-(1+\mu ) K t}
\frac{1-A e^{(1+\mu )Kt}}{1-A} \cdot 
\frac{1+\mu A e^{(1+\mu )Kt}}{1+\mu A}\notag\\
&&+\int_0^t d\tau
\left(-1+\mu K\alpha\frac{1-Ae^{(1+\mu )Kt}}{1+\mu Ae^{(1+\mu )Kt}}\right) 
e^{-(1+\mu )K(t-\tau)}\frac{1-A e^{(1+\mu )Kt}}{1-Ae^{(1+\mu )K\tau}}
\cdot\frac{1+\mu A e^{(1+\mu )Kt}}{1+\mu Ae^{(1+\mu )K\tau}}\ .
\end{eqnarray}
The results for the typical trajectories leading to a solution after
some given final time $t_f$ are presented in Fig. 
(\ref{fig:3satendens}). They show the same qualitative behavior
like $K$-XOR-SAT with a slightly slower convergence towards the
equilibrium due to the reduced exponential factor $e^{-(1+\mu)Kt}$.
Also the action calculated for the trajectories shows a similar
behavior like for $K$-XOR-SAT, cf. Fig. \ref{fig:3satwirk}. The
exponentially dominant contribution to the typical solution time is 
again given by $t_{sol} \sim \lim_{t_f\to\infty} e^{NS(t_f)}$.

\begin{figure}[htbp]
  \begin{center}
    \includegraphics[clip, height=7cm]{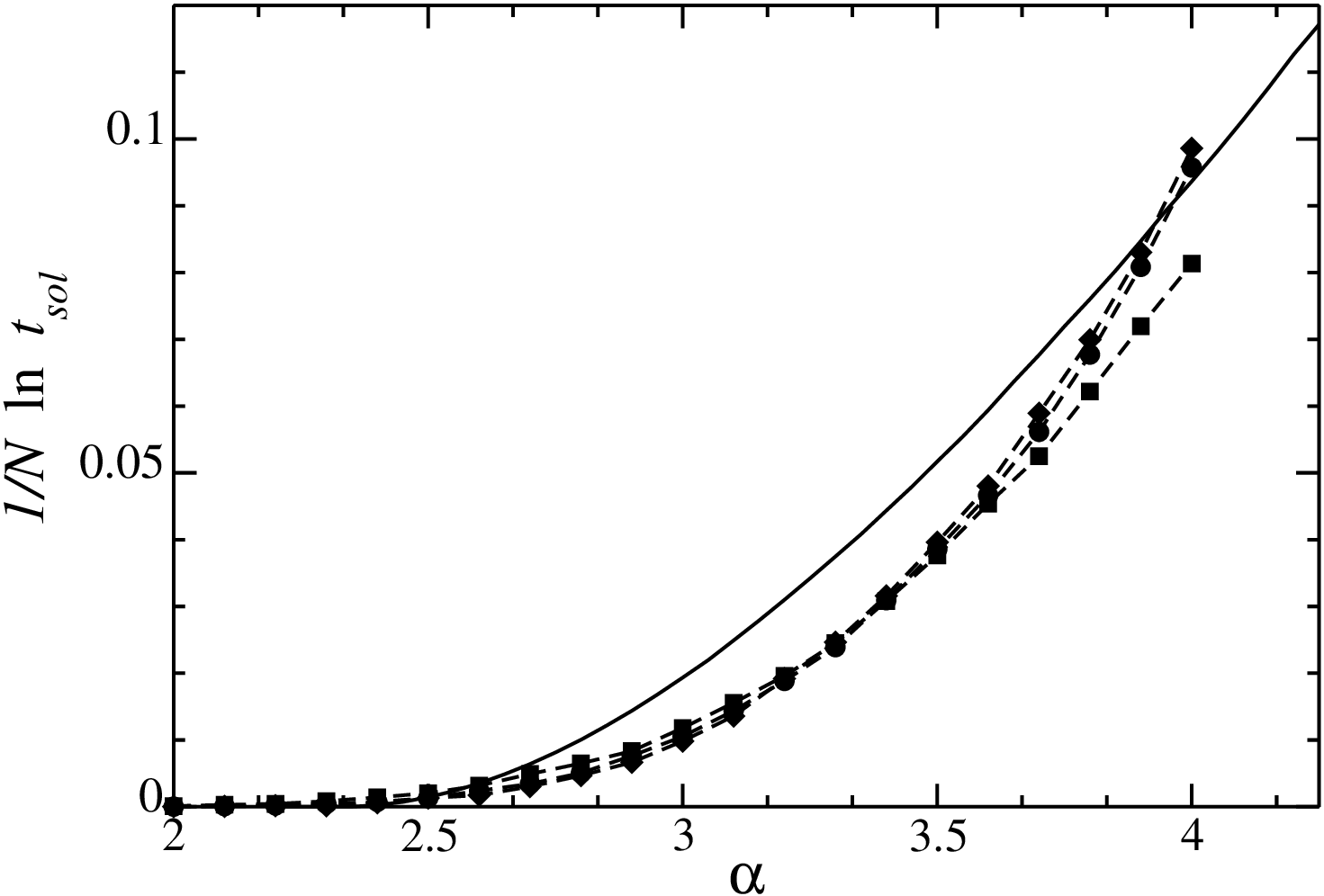}
    \caption{3-SAT: 
      Solution time $t_{sol}$ for Sch\"oning's algorithm (only walk
      steps, random restarts after $t_fN=3N$ steps) measured as the
      number of restarts, as a function of $\alpha$. The analytical
      result is given by the full line. Numerical data for
      $N=30,50,70$ (squares, dots, diamonds) this time cross the
      analytical prediction. Note that the solution times are smaller
      than for 3-XOR-SAT.}
    \label{fig:3sattime}
  \end{center}
\end{figure}

In Fig. \ref{fig:3sattime} we finally compare the predicted typical
solution time with numerical simulations. Close to the dynamical
threshold, the numerical running times are much smaller, which can be
explained already by the fact that the Poissonian approximation
under-estimates $\alpha_d$. For larger $\alpha$, the numerical data
cross the analytical approximation, but both stay well below
Sch\"oning's bound. This is to be expected, since there is an
exponential number of possible solutions, while Sch\"oning assumes only the
existence of a single one. Note that the solution times are exponentially
smaller for 3-SAT than for random 3-XOR-SAT.


\section{Conclusion and outlook}
\label{sec:conclusion}

In this paper, we have presented an approximate analytical approach to
describe the dynamical behavior of a class of stochastic local search
algorithms applied to random $K$-satisfiability and
$K$-XOR-satisfiability problems. We have seen that there are two
distinct dynamical phases:
\begin{itemize}
\item For clause-to-variable ratio $\alpha<\alpha_d$ (with $\alpha_d$
  being algorithm- and problem-dependent), the algorithm is able to
  solve almost all instances in linear time. In this regime, the
  dynamics was studied using a simple rate-equation approach which was
  able to capture the most important features of the average
  trajectory taken by the system under the action of the algorithm.
\item For $\alpha>\alpha_d$, typical solution times were found to
  scale exponentially with the system size given by the number of
  variables $N$. This behavior could be understood analytically using
  a functional-integral approach to evaluate the probability of large
  deviations from the typical trajectory. We found the following
  behavior: The system equilibrates very fast to a non-zero plateau in 
  the number of unsatisfied clauses. Then the system only fluctuates
  around this plateau. This goes on until an exponentially improbable
  macroscopic fluctuation towards one of the solutions appear, and the
  algorithm stops. The small probability of these fluctuations
  explains the exponentially large waiting times until a satisfying
  assignment is reached.
\end{itemize}
For the exponential-time regime, only a Poissonian approximation was
used. In principle it would be possible to go beyond this ansatz using
the full distribution $p_t(s,u)$ of vertices with $s$ satisfied and
$u$ unsatisfied clauses. Following the same scheme  as in the
Poissonian approach, we reach a system of first-order
differential equations for all $p_t(s,u)$ and their conjugate
parameters $\kappa_t(s,u)$. Being non-linear, it is far from obvious
how to construct an analytical solution. But also the numerical
integration of these equations is a hard problem: For the $p_t(s,u)$
there are initial and final conditions, whereas the $\kappa_t(s,u)$
have no boundary condition at all. The question if it is possible to
follow this improved approach is still under investigation.

Another possible extension of this work concerns the application of
different heuristics like GSAT which was discussed in the second
section. The analytical approach can serve as a basis for evaluating  
the relative performance of different heuristics and, as a consequence
of the insight gained, also as a step towards a systematic improvement
of stochastic local search.

A third point which remains open is the question in how far the
solution space structure influences the performance of walk-SAT. As
discussed in the beginning of the paper, random $K$-SAT and random
$K$-XOR-SAT undergo a clustering transition deep inside the
satisfiable phase.  Below this transition, all solutions are collected
in one huge cluster, above, an exponential number of such clusters
exists. The clustering transition is also connected to a proliferation
of metastable states which are expected to cause problems for any local
algorithm. However, in our approach to the walk-SAT dynamics, we do
not see any sign of a direct impact of this transition on the
performance of the algorithms under consideration. The onset of
exponential solution times is found to be inside the unclustered
phase. It thus remains an open problem whether the clustering
transition can be approached by using improved heuristic criteria.

{\bf Acknowledgments:} We are grateful to R. Zecchina for helpful
discussions. We also thank R. Monasson and G. Semerjian for
communicating their results \cite{MoSe} prior to publication. WB and
AKH obtained financial support from the DFG (Deutsche
Forschungsgemeinschaft) under grant Zi 209/6-1.  AKH was also partly
funded by the VolkswagenStiftung within the program "Nachwuchsgruppen
an Universit\"aten".


\begin{thebibliography}{99}
\bibitem{note} As we will see later on, the case $q<1$ is not well
  described by the independent-neighbor approximation. This is also
  true for the stronger Poissonian approximation.
\end{thebibliography}
\end{document}